\documentclass[onecolumn,showpacs,preprintnumbers,amsmath,amssymb,12pt]{revtex4}
\usepackage{graphicx}% Include figure files
\usepackage{dcolumn}% Align table columns on decimal point
\usepackage{bm}% bold math

\newcommand{\be}{\begin{equation}}
\newcommand{\bea}{\begin{eqnarray}}
\newcommand{\ee}{\end{equation}}
\newcommand{\eea}{\end{eqnarray}}
\begin{document}

\title{Reading Sequences of Interspike Intervals in Biological Neural Circuits}% Force line breaks with \\

\author{Henry D. I. Abarbanel}
\affiliation{Department of Physics\\
and\\
Marine Physical Laboratory (Scripps Institution of Oceanography)}
\altaffiliation[Also at]{ Institute for Nonlinear Science}
\altaffiliation{University of California, San Diego; La Jolla, CA
92093-0402  USA}

\author{Sachin S. Talathi}
\affiliation{Department of Physics\\ and\\ Institute for Nonlinear
Science} \altaffiliation{University of California, San Diego; La
Jolla, CA 92093-0402  USA}
\email{talathi@physics.ucsd.edu}

\date{\today}
\clearpage
\begin{abstract}

Sensory systems pass information about an animal's environment to
higher nervous system units through sequences of action
potentials. When these action potentials have essentially
equivalent waveforms, all information is contained in the
interspike intervals (ISIs) of the spike sequence. We address the
question: How do neural circuits recognize and read these ISI
sequences ?

Our answer is given in terms of a biologically inspired neural
circuit that we construct using biologically realistic neurons.
The essential ingredients of the ISI Reading Unit (IRU) are (i) a
tunable time delay circuit modelled after one found in the
anterior forebrain pathway of the birdsong system and (ii) a
recently observed rule for inhibitory synaptic plasticity. We
present a circuit that can both learn the ISIs of a training
sequence using inhibitory synaptic plasticity and then recognize
the same ISI sequence when it is presented on subsequent
occasions. We investigate the ability of this IRU to learn in the
presence of two kinds of noise: jitter in the time of each spike
and random spikes occurring in the ideal spike sequence. We also
discuss how the circuit can be detuned by removing the selected
ISI sequence and replacing it by an ISI sequence with ISIs drawn
from a probability distribution.

We have investigated realizations of the time delay circuit using
Hodgkin-Huxley conductance based neurons connected by realistic
excitatory and inhibitory synapses. Our models for the time delay
circuit are tunable from about 10 ms to 100 ms allowing one to
learn and recognize ISI sequences within that range of ISIs. ISIs
down to a few ms and longer than 100 ms are possible with other
intrinsic and synaptic currents in the component neurons.

\end{abstract}

\pacs{Valid PACS appear here}

\maketitle

\clearpage

\noindent
\section{Introduction}
Sensory systems transform environmental signals into a format
composed of essentially identical action potentials.
 These are sent for further processing to other areas of a central nervous system.
When the action potentials or spikes are comprised of identical
waveforms all information about the environment is contained in
the spike arrival times~\cite{fano}. The spike train can also be
characterized in terms of interspike interval times for a given
spiking sequence.

There are many examples of sensitive stimulus-response properties characterizing
how neurons respond to specific stimuli. These include whisker-selective neural
response in barrel
cortex~\cite{Welker,Aara} of rats and motion
sensitive cells in the visual cortical areas of primates  ~\cite{Sugase,Buracas}.

One striking example is the selective auditory response of neurons
in the songbird telencephalic nucleus HVC (proper name)
~\cite{lewiiki,Margoliash1,Margoliash2,Coleman1}. (Good
introductions to the birdsong system may be found in the paper by
Brenowitz, Margoliash, and Nordeen~\cite{bren97} and the review by
Brainard and Doupe~\cite{brain02}.) Projection neurons within HVC
fire sparse bursts of spikes when presented with auditory playback
of the bird's own song (BOS) and are quite unresponsive to other
auditory inputs. The auditory nucleus NIf (interfacial nucleus of
nidopallium), through which auditory signals reach
HVC~\cite{Janata,Carr,Coleman1,raskin}, also strongly responds to
BOS in addition to responding to a broad range of other auditory
stimuli. NIf projects to HVC, and the similarity of NIf responses
to auditory input and the subthreshold activity in HVC neurons
suggests that NIf could be acting as a filter for BOS ,
preferentially passing that important signal on to HVC. It was
these examples from birdsong that led us to address the ISI
reading problem we consider here.

The representation of important neural information in ISI
sequences suggests the presence of a biological neural network,
perhaps widely used across species, capable of accurately
recognizing specific ISI sequences. In this paper we develop such
a network and demonstrate how biologically realistic neurons and
synapses can be used to construct and train such a network to
recognize specific ISI sequences. We call the resulting network an
ISI Reading Unit (IRU).

Key to the functioning of an IRU are two biological processes:
\begin{enumerate}
\item  a time delay unit which, on receiving a spike at time $t_{0}$, produces an output
spike at time $t_{0}+\tau(R)$, where R is a dimensionless
parameter characterizing synaptic strength that can be used to
tune the time delay $\tau(R)$;
\item a method for tuning the time delays in the IRU $\tau(R)$ using synaptic
plasticity of inhibitory synapses as observed recently~\cite{Haas}.
\end{enumerate}

Time delay circuits, thought of primarily as an abstract idea
rather than as a particular biological circuit realization, have
been considered before~\cite{Buono,Mauk,Ivry}. An exception to the
descriptive modelling of neural time telling processes is the work
of Buonomano~\cite{dean} which studies a two neuron model which
can be tuned to respond to time delays. Buonomano identifies
synaptic changes as the tuning mechanism that might underlie
detection of time intervals. His model relies on a balance between
excitatory and inhibitory synaptic strengths. His construction
might provide an alternative to the time delay unit structure we
explore here, but we do not present an analysis of his results
here.

As discussed by these
authors circuits for telling time more or less divide into three categories:
\begin{itemize}
\item time delays along pieces of axon resulting
in delays as short as a few microseconds and found in detection
circuits for inter aural time differences~\cite{knudsen};

\item time delays of order hours or days connected with circadian rhythms, and

\item time delays of tens to hundreds of milliseconds associated
 with cortical and other neural processing.
\end{itemize}

Our realization of a time delay circuit addresses this third
category using ideas from an observed neural circuit in the
birdsong system.

In investigating time differences between signals propagating from
the birdsong nucleus HVC directly to the pre-motor nucleus RA and
the same signal propagating to RA around the neural loop known as
the anterior forebrain pathway (AFP), Kimpo, Theunissen, and
Doupe~\cite{kimpo} reported a remarkable precision of the time
difference between these pathways of 50 $\pm$ 10 ms across many
songbirds and many trials.

In investigating models of this phenomenon and its implications,
along with excitatory synaptic plasticity at the HVC $\to$ RA
junction (robust nucleus of acropallium) (see inset figure
~\ref{threeneuron}), we~\cite{Abar1} constructed a circuit of
neurons based on detailed electrophysiological measurements by
Perkel and his colleagues~\cite{perkel1,perkel2}, in each of the
three nuclei of the AFP. This circuit demonstrated a tunable time
delay adjusted by the strength of inhibition of synapse from the
nucleus Area X to the nucleus DLM. The precise value of the time
delay in the birdsong circuit was attributed to a fixed point in
the overall dynamics including excitatory synaptic plasticity at
the HVC $\to$ RA junction. This investigation suggested a general
form of time delay circuit that could be tuned by changing the
strength of an inhibitory synaptic connection. We develop that
idea here.

Using these biologically motivated ingredients, we have built a
simplified neural time-delay circuit and show here how it can be
tuned to produce time delays within the range of about 10 ms to
100 ms or so. We demonstrate how to train such a circuit using
given specific ISI sequences, and then show how the trained
circuit robustly recognizes the desired ISI sequence. Recognition
is implemented here using a detection circuit that fires an action
potential when two input spikes arrive within a short temporal
window of $\delta$ ms, (taken here to be 1 ms), and responds with
subthreshold activity otherwise.

The overall IRU circuit made up of a subcircuit of time delay
units operates by producing a replica of the given ISI sequence.
It then uses inhibitory synaptic plasticity to adjust the delays
in a sequence of time-delay sub circuits to match the ISIs in the
input sequences within the resolution threshold of $\delta$ ms.
Success in this matching is seen in the spiking activity of the
detection circuit. The IRU circuit is thus a candidate for how
biological networks can accurately select particular environmental
signals, potentially usable for further processing for decision
making and required functionality, by keying on the representation
of environmental signals as a specific spike sequence.

We first discuss the construction of the time delay circuit
beginning with the design of a smallest possible biologically
feasible neural circuit consisting of two neurons and then explain
the mechanism of the three neuron time delay circuit abstracted
from the birdsong system. We then address the issue of training
this delay circuit using synaptic plasticity of inhibitory
synapses to detect an input ISI sequence. We then proceed to
develop our full ISI Reading Unit (IRU) and show how it can be
used to recognize a specific ISI sequence on which it has been
trained. We show that the circuit can be trained robustly in two
types of noisy environments: (1) When there is random jitter on
the ISIs in the input sequence, and (2) when extra spikes are
randomly inserted into the ISI sequence. The latter represents, in
a realistic biological setting, the presence of other spikes
associated with additional activity in the system as a whole. We
also explore how an IRU can be detuned when presentation of the
selected ISI sequence is replaced by a random ISI sequence.

\section{A Biological Time-Delay Circuit}

A two neuron time delay circuit is motivated by construction of a
single neuron model of type I ~\cite{Ermen}, that has property of
being able to produce spikes at very low frequencies. It can be
shown that ~\cite{Izhi} for neurons of type I, the frequency of
spiking as function of constant input current $I_{IN}$, obeys the
following relationship,$$f=C\sqrt{I_{IN}-I_{0}}$$, where $I_{0}$,
is the spiking threshold and C is the scaling constant, which is
function of model parameters. This frequency current relationship
is characteristic of saddle node bifurcation of neuron to spiking
regime, typically observed in Type 1 neurons. A neuron model
developed by Traub ~\cite{Traub} consisting of transient sodium
channels, delayed rectifier potassium channel and a leak channel,
described by following set of dynamical equations, has frequency
current relationship of Type I neuron behavior.
\begin{eqnarray*}
C_{M}\frac{dV_{I}(t)}{dt}&=&g_{Na}m(t)^{3}h(t)(V_{Na}-V_{I}(t)) \nonumber \\
&+&g_{K}n(t)^{4}(V_{K}-V_{I}(t))+g_{L}(V_{L}-V_{I}(t)) \nonumber \\
&+&I_{IN} \nonumber \\
\end{eqnarray*}
where $V_{I}$ is the membrane voltage of neuron. $C_{M}$, is the
membrane capacitance, $g_{Na}$, $g_{K}$ and $g_{L}$ are maximal
membrane conductance of the transient sodium channel, the delay
rectifier potassium channel and the leak channel. $m(t)$ and
$n(t)$ are the activation gating variables, which open up when
membrane voltage increases and $h(t)$ represents the inactivation
gating variable which closes on increasing membrane potential.
$V_{Na}$, $V_{K}$ and $V_{L}$ represent the reversal potential for
sodium, potassium and the leak channels respectively. The
particular values for the parameters used in our equation are
given in the appendix.

\begin{figure*}[ht!]
\begin{center}
\includegraphics[width=7.in,scale=1,angle=0]{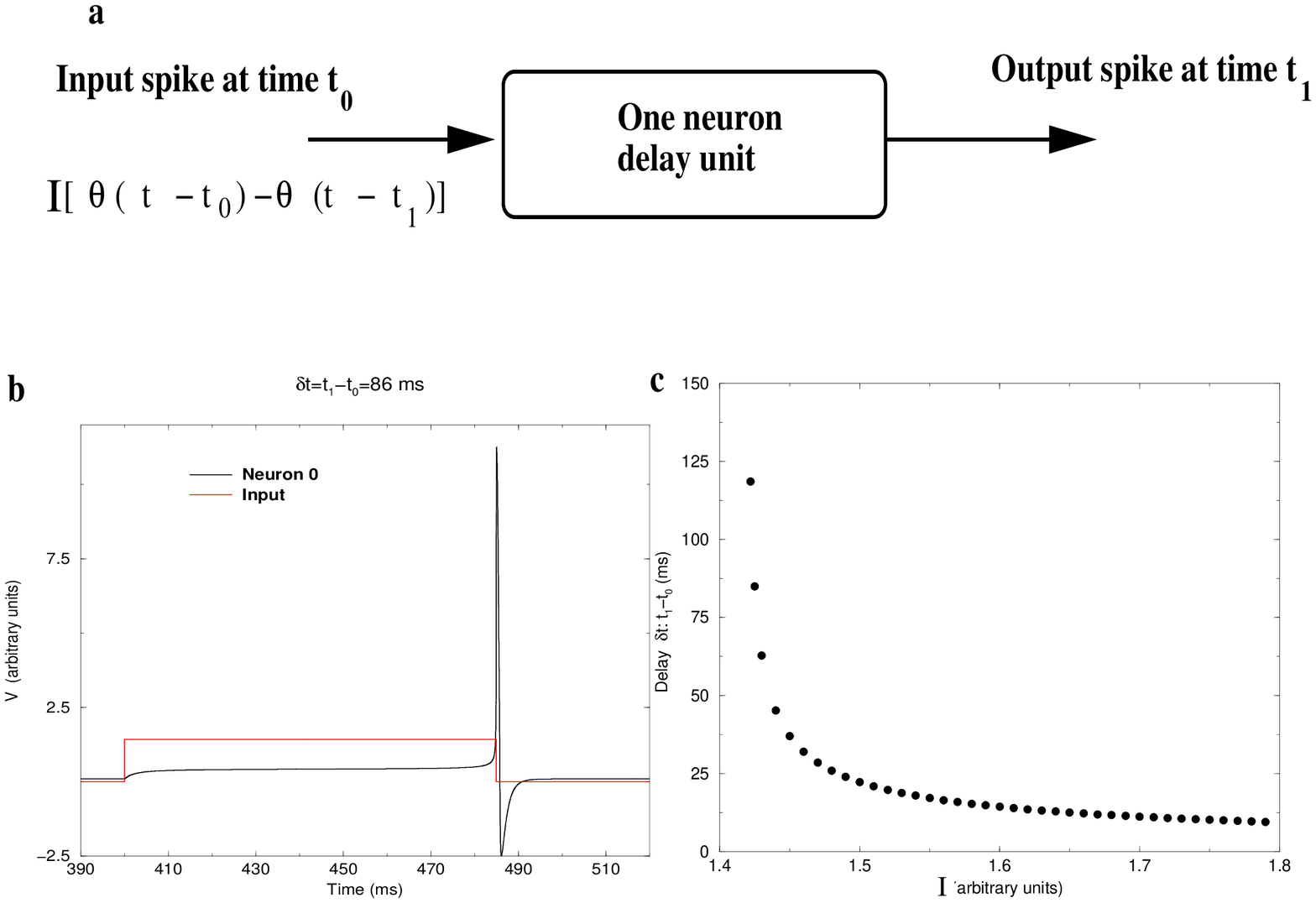}
\caption{(a) Schematic of the one neuron delay unit model. Input
to the neuron is a step current starting at time $t_{0}$ of the
input spike and lasting until the first spike from the neuron. The
intrinsic dynamics of the neuron to the spiking mode is through a
saddle node bifurcation, typical of Type I neurons, that governs
the delay produced by the neuron. (b) Scaled output from the delay
unit, in response to step input current is shown. In this
particular case, the neuron produces a delayed spike after about
87 ms. (c) Variation of the delay produced by the neuron as
function of the strength of the input step current.
\label{oneneuron}}
\end{center}
\end{figure*}

\begin{figure*}[ht!]
\begin{center}

\includegraphics[width=7.in,scale=1,angle=0]{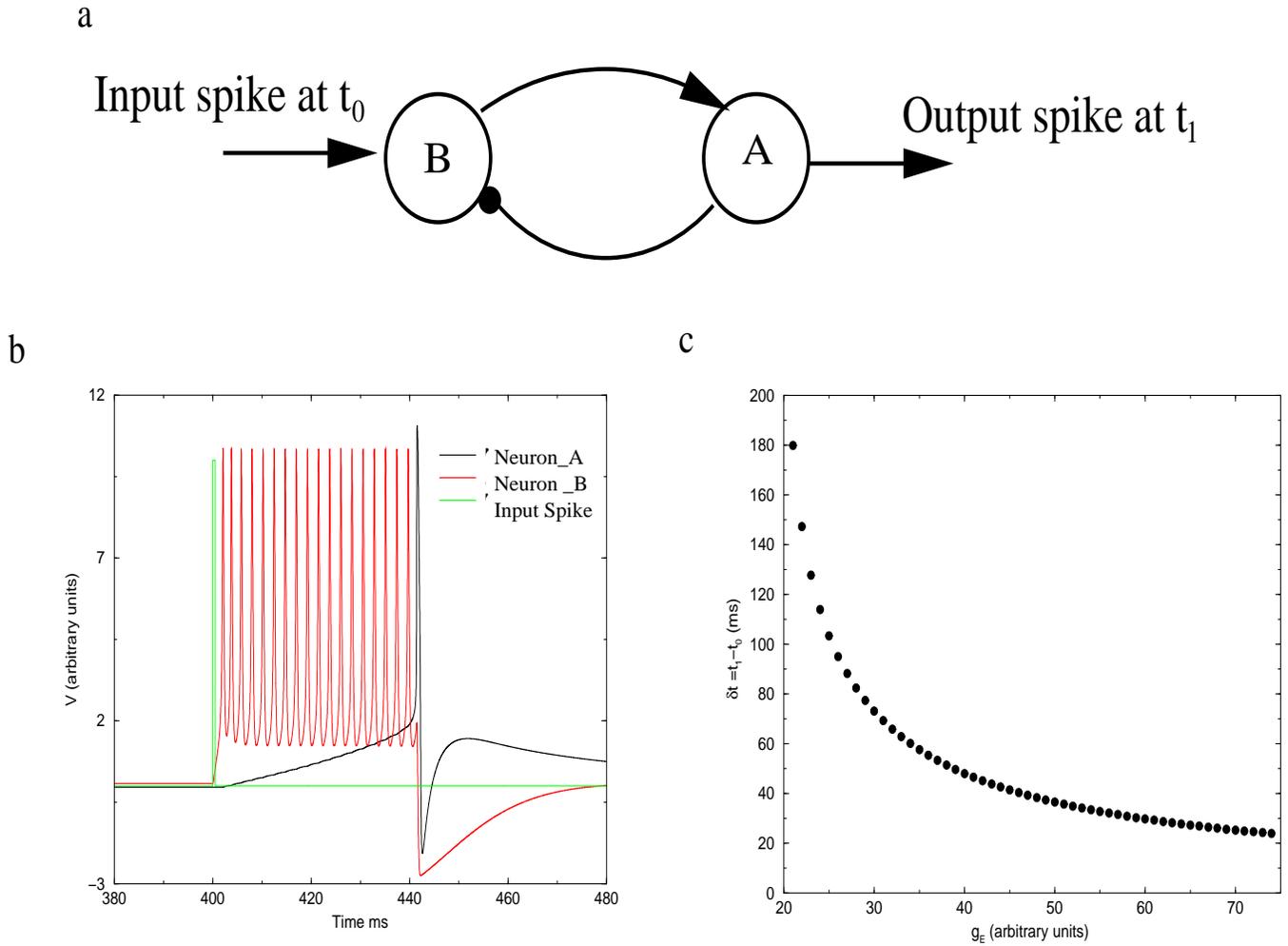}
\caption{Schematic of a two neuron delay unit model. The input
spike arrives at neuron B at time $t_{0}$, which pushes this
bistable neuron into a spiking regime, raising the neuron A
membrane voltage towards spiking threshold, until it eventually
spikes. The spike from neuron A, pushes neuron B back into a
stable resting state. (b) The membrane voltage of neuron A and
neuron B, in response to an input spike at time $t_{0}=400$ ms.
For this particular case, neuron A fires after a delay of around
42 ms. The delay produced is governed by the strength of the
excitatory synaptic connection from neuron B to neuron A. (c) Plot
of delay as a function of the strength of the excitatory synaptic
input from neuron B to neuron A. \label{twoneuron}}
\end{center}
\end{figure*}

Using this neuron model, we construct a single neuron time delay
circuit as shown in Figure ~\ref{oneneuron}a. The input to the
neuron is $I\left[\Theta(t-t_{0})-\Theta(t-t_{1})\right]$, where
$\Theta(x)=0$ when $x\le 0$ and $\Theta(x)=1$ when $x>0$. $t_{0}$,
is the time of an input spike to the neuron and $t_{1}$, is the
time of first output spike out of the neuron. The period of
spiking $T=(1000/f)$ ms, for type I neurons obeys
$T=1000/C\sqrt{I_{IN}-I_{0}}$, i.e., $T\rightarrow \infty$, as
$I_{IN}\rightarrow I_{0}$. Therefore, using such a type I neuron
model, with step input current, for the duration of first spike,
we can construct a timedelay unit to obtain various time delays as
a function of strength of $I_{IN}$. In Figure ~\ref{oneneuron}a,
we show the delay $\tau=t_{1}-t_{0}$, from this one neuron model
as a function of $I_{IN}$. A biologically feasible smallest
possible model for time delay with two neurons using an extension
of above ideas from our one neuron model is shown in Figure
~\ref{twoneuron}a. The circuit is composed of two neurons, neuron
B which exhibits the property of bistability, of coexistent
resting state and spiking state, which is observed in many a
neuron models ~\cite{Guttman} and neuron A, which is of type I as
discussed above. The input to this delay circuit is a spike
occurring at time $t_{0}$ . The output of the model is a single
spike at time $t_{1}$ from neuron A, and thus $\tau=t_{1}-t_{0}$.
This can be changed as a function of the maximal conductance of
the excitatory synaptic connection from neuron B to neuron A. A
single input spike into neuron B moves it into its spiking regime,
providing enough depolarizing current into neuron A until it fires
a spike. In order for spiking neuron B to provide enough
depolarization to neuron A for it to eventually fire, the firing
frequency of neuron B should be greater than the decay time of the
excitatory synaptic connection from B to A. This necessitates the
use of slow NMDA type excitatory synaptic connection from neuron B
to A. A spike from neuron A, then provides a hyperpolarizing input
to neuron B pushing it back into its rest state. In Figure
~\ref{twoneuron}c we show the plot of the time delay of this model
as function of the maximal conductance of the excitatory synaptic
input to neuron A from neuron B. An interesting feature of this
two neuron time delay model is that, $\frac{d\tau}{dR}<0$. We will
see in a later section on training the IRU to detect a given spike
sequence, using observed learning rules of synaptic plasticity,
the necessity of having $\frac{d\tau}{dR}>0$. This means that, the
use of this two neuron unit delay model is infeasible for
construction of IRU.

\begin{figure*}[ht!]
\begin{center}
\includegraphics[width=7in,scale=1,angle=0]{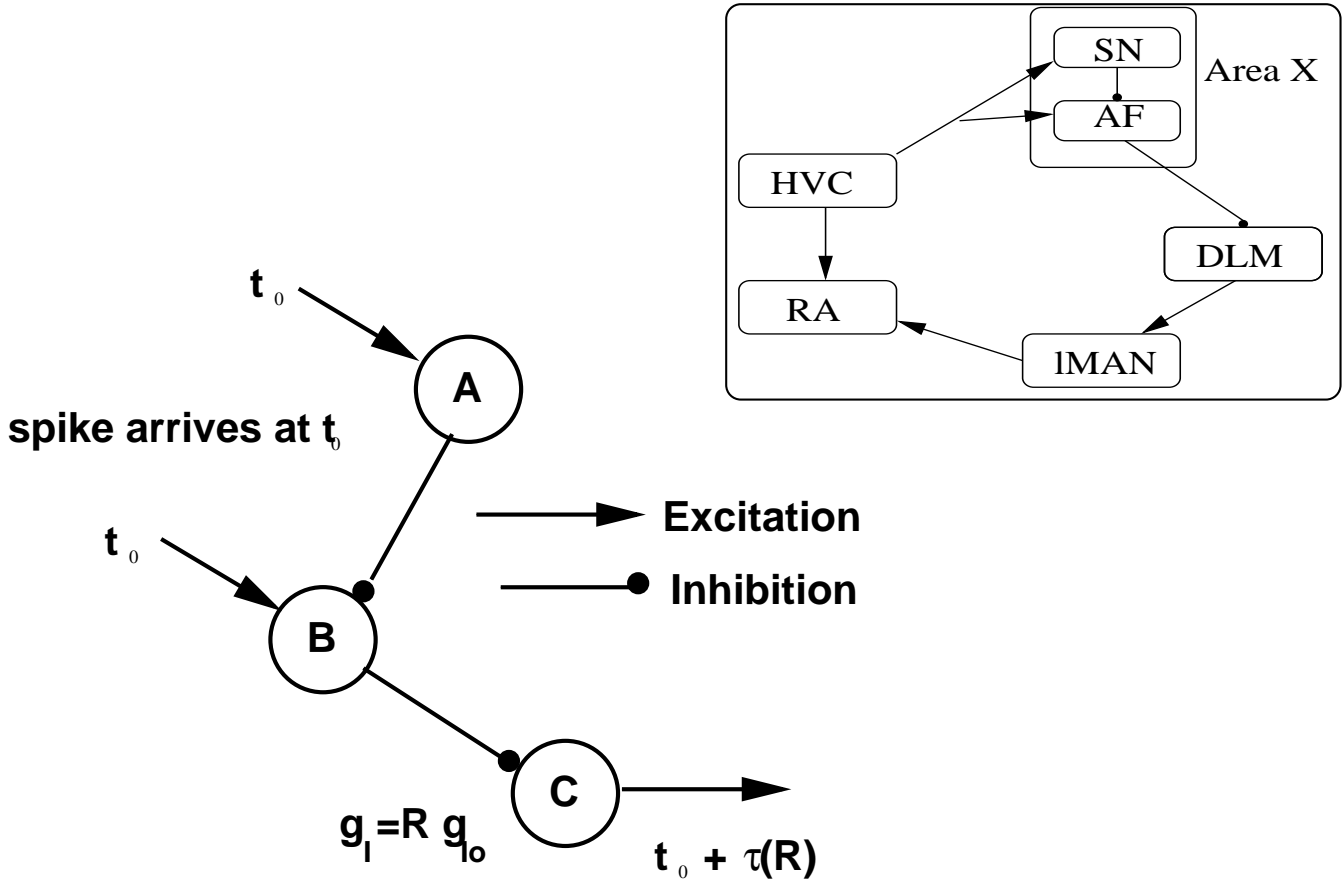}
\caption{Schematic diagram of the time-delay unit used in the IRU
circuit. This is abstracted from a time delay network in the
anterior forebrain pathway of the birdsong system as shown in the
inset above. Th einset shows the AFP loop (Area X, DLM and lMAN)
from the birdsong system that suggested our three neuron timedelay
unit. Absent any input spikes, neuron A is at rest, neuron B
oscillates periodically, and neuron C oscillates around its rest
potential driven by periodic inhibitory input from neuron B. When
an input spike arrives at neuron A and at neuron B at time
$t_{0}$, neuron A fires an action potential and neuron B has the
phase of its oscillation reset to be in synchrony with the time of
arrival $t_0$ of the spike. The action potential in neuron A
inhibits neuron B, and this releases neuron C to rise to its
spiking threshold a time $\tau(R)$ later. R is the dimensionless
scale of the $B \to C$ inhibition. Within a broad range for R,
neuron C will fire a single spike at a time $t_0 + \tau(R)$. The
value of the conductance for the B $\to$ C inhibitory synapse is
$g_I = Rg_{I0}$, with $g_{I0}$ a baseline conductance..
\label{threeneuron}}
\end{center}

\end{figure*}

The simplest possible biologically feasible three neuron model for
time delay circuit is motivated by the anterior forebrain pathway
(AFP) loop in the song bird brain (Figure \ref{threeneuron}
inset). The AFP of songbirds is comprised of three nucles, the
area X nucles, the DLM nucles and the lMAN nucles, each having a
few times 10,000 neurons~\cite{brain02}. The input to the AFP  is
via sparse burst of spikes from nucles HVC , entering via Area X,
and the output signal of the AFP is from lMAN leaving the AFP to
innervate the RA nucles. Within Area X two distinct neurons, spiny
neurons (SN) and aspiny fast firing (AF) neurons, receive direct
innervation from HVC~\cite{perkel1, Perkelrecent}. In the absence
of signals from HVC, the SNs are at rest while the AF neurons are
oscillating at about 20 Hz. The SNs inhibit the AF neurons, and
these, in turn, inhibit neurons in DLM, a thalamic nucleus in the
AFP~\cite{Farries}. The DLM neurons receiving this input from Area
X are below threshold for action potential production while the AF
neuron oscillates, but when the AF $\rightarrow$ DLM inhibition is
released, the DLM neurons fire an action potential. This is
propagated to lMAN, and then transmitted to RA. The time around
this path differs from direct HVC $\rightarrow$ RA innervation by
$50 \pm 10\,$ ms ~\cite{kimpo}.

In our modelling of this observation~\cite{Abar1}, treating each
nucleus as a coherent action potential generating device, we found
that LMAN played an unessential role in determining the time delay
around the AFP while the strength of the AF $\rightarrow$ DLM
inhibition could tune the time delay over a few tens of
milliseconds.

From these observations, we have constructed a biologically
feasible time delay circuit comprised of three neural units and
two inhibitory synapses with a tunable synaptic strength.

The time delay circuit is displayed in Figure~\ref{threeneuron}.
Neuron A (similar to the SN in Area X) receives an excitatory
input signal from some source. It is at rest when the source is
quiet, and when activated it inhibits neuron B. Neuron B receives
an excitatory input from the same source. It oscillates
periodically when there is no input from the source. Neuron B
inhibits neuron C. Neuron C produces periodic spiking in the
absence of inhibition from neuron B.

In this paper each of the neurons A, B, and C is represented by a
simple Hodgkin-Huxley (HH) conductance based model similar to the
Traub model we discussed earlier with sodium, potassium, and leak
currents as well as an injected DC current to set the spiking
threshold. A more detailed neuron model for neuron C could include
hyperpolarization activated $I_{h}$ channels and low threshold
calcium $I_{T}$ channels, which facilitate post inhibitory rebound
spikes~\cite{Abar1}. Indeed, in the DLM neuron of the birdsong AFP
this mechanism leads to calcium spikes as the output of ``neuron
C."

When the inhibition from neuron B to neuron C is released by the
signal from neuron A to neuron B, neuron C rebounds and produces
an action potential some time later. This is due to its intrinsic
stable spiking of neuron C in the absence of any inhibition from
neuron B.

This time delay is dependent on the strength of the
B$\rightarrow$C inhibition, as the stronger that is set the
further below threshold neuron C is driven and the further it must
rise in membrane voltage to reach the action potential threshold.
This means the larger the B$\rightarrow$C inhibition, the longer
the time delay produced by the circuit. Other parameters in the
circuit, such as the membrane time constants, also set the scale
of the overall time delay.

The direct excitation of neuron B by the signal source is
critical. It serves to reset the phase of the neuron B
oscillation, as a result of which the spike from neuron C is
measured with respect to the input signal and thus makes the
timing of the circuit precise relative to the arrival of the
initiating spike. Absent this excitation to neuron B, the phase of
its oscillation is uncorrelated with the arrival time of a signal
from the source, and the time delay of the circuit varies over the
period of oscillation of neuron B. This is not a desirable
outcome, nor is it the way the AFP circuit appears to work.

We have constructed this circuit using standard Hodgkin-Huxley
(HH) conductance based neurons and realistic synaptic connections.
The dynamical equation for the three neurons shown in figure
~\ref{threeneuron} are.
\begin{eqnarray*}
C_{M}\frac{dV_{i}(t)}{dt}&=&g_{Na}m(t,V_{i}(t))^{3}h(t,V_{i}(t))(V_{Na}-V_{i}(t)) \nonumber \\
&+&g_{K}n(t,V_{i}(t))^{4}(V_{K}-V_{i}(t))+g_{L}(V_{L}-V_{i}(t)) \nonumber \\
&+&g^{I}_{ij}S_{I}(t)(V_{revI}-V_{i}(t)) + I^{syn}_{i}(t)+
I^{DC}_{i},
\end{eqnarray*}
where (i,j)=[A, B, C]. The membrane capacitance is $C_{M}$, and
$V_{Na},V_{K},V_{L}$, and $V_{revI}$ are reversal potentials for
the sodium, potassium, leak, and inhibitory synaptic connection
respectively. $m(t), h(t)$, and $n(t)$ are the usual activation
and inactivation dynamical variables; the equations for these are
given in an appendix. $g_{Na},g_{K},g_{L}$ and $g_{E}$ are the
maximal conductances of sodium, potassium and leak channels and
excitatory connections respectively. $I^{DC}_{i}$ is the DC
current into the A, B or C neuron. These are selected such that
neuron A is resting at -63.74 mV in absence of any synaptic input,
neuron B is spiking at around 20 Hz, and neuron C is also spiking
at around 20 Hz, in absence of any synaptic inputs.
$I^{syn}_{i}=g^{E}_{i}S_{E}(t)(V_{revE}-V_{i}(t))$ is the synaptic
input to the delay circuit at neuron A and B, from the signal
source at time $t_{0}$, where $g^{E}=(g_{EA},g_{EB},0)$. The
inhibitory synaptic strengths, $g^{I}_{ij}$ in the delay circuit
are $g_{BA}=R^{0}g_{I}$ and $g_{CB}=Rg_{I}$.

The dimensionless factors R and $R^{0}$, set the strength of
$B\rightarrow C$  and $A\rightarrow B$ inhibitory connections
respectively, relative to baseline strength $g_{I}$

$S_{E}(t)$ represents the fraction of neurotransmitter, docked on
the postsynaptic cell receptors as a function of time. It varies
between 0 and 1 and has two time constants: one for the docking
time of the neurotransmitter and one for its release time. It
satisfies the dynamical equation:
\begin{eqnarray*}
\frac{dS_{E}(t)}{dt}&=&\frac{S_{0}(V_{pre}(t))-S_{E}(t)}{\tau_{E}(S_{1E}-S_{0}(V_{pre}(t)))}.
\end{eqnarray*}

The docking time constant for the neurotransmitter is
$\tau_{E}(S_{1E}-1)$, while the undocking time is
$\tau_{E}S_{1E}$. The function $S_{0}(V_{pre}(t))$, is 0 when
$V_{pre}(t)=0$, representing no spike input from presynaptic
terminal, and is 1 when $V_{pre}(t) > 0$, representing spike input
from the presynaptic terminal. For neurons A and B the presynaptic
voltage is given by the incoming spike or burst of spikes arriving
from some source at time $t_{0}$ (See figure ~\ref{threeneuron}).
For our excitatory synapses we take $\tau_{E}=1\,$ms and
$S_{1E}=1.5$, for a docking time of 0.5 ms and an undocking time
of 1.5 ms. These times are characteristic of AMPA excitatory
synapses.

Similarly $S_{I}(t)$ represents the percentage of
neurotransmitter, docked on the postsynaptic cell as function of
time. It satisfies the following equation

\begin{eqnarray*}
\frac{dS_{I}(t)}{dt}&=&\frac{S_{0}(V_{pre}(t))-S_{I}(t)}{\tau_{I}(S_{1I}-S_{0}(V_{pre}(t)))}
\end{eqnarray*}
where we select $\tau_{I}=1.2\,$ ms and $S_{1I}=4.6$ for docking
time of 4.32 ms and undocking time of 5.52 ms. The range of time
delays produced by the three neuron delay circuit, depends on the
docking and undocking times of this synapse. For the values given
above we get the the delay curve as shown in figure ~\ref{delay}.
The inhibitory reversal potential $V_{revI}$ is chosen as -80mV.
The voltage presynaptic to neuron B is $V_A(t)$ and when A spikes
$V_{pre}(t)>0$ in the equation for $S_I(t)$ above.

The various parameters as well as the dynamical equations for the
common activation and inactivation variables are presented in the
appendix.

In Figures ~\ref{membrane1}a and ~\ref{membrane1}b we present
examples of the response of the time delay circuit just described.
In Figure ~\ref{membrane1} a single spike is presented to neurons
A and B at $t_{0}=500.0\,$ms. For R = 0.7 the resulting time delay
produced by the delay unit is 43.68 ms. In this figure we show the
membrane voltage of neurons A and C. In Figure ~\ref{membrane1}b
we also show the membrane voltage of neuron B. It is clear in
Figure ~\ref{membrane1}b that the oscillations of neuron B are
reset by the incoming signal, and the action potential generated
at neuron C is a result of its internal dynamics, not of the
period of oscillation of neuron B. The variation of the time delay
with the strength of the B$\rightarrow$C inhibitory strength R is
shown in Figure ~\ref{delay} for the particular set of parameters
listed in the appendix. The important point here is that
$\frac{d\tau(R)}{dR}>0$. For $R<R_{L}$, the inhibition on neuron C
from neuron B, is not strong enough and the spike from neuron C is
no longer correlated to input spike to the delay circuit. For
$R>R_{U}$, the inhibition is so strong that neuron C does not
produce rebound spike at all. As noted, one can, by changing the
membrane capacitance and the strengths of the various maximal
conductances and time course of inhibitory synaptic connections,
place the variation of $\tau(R)$ near 10 ms or near 100 ms.

\section{An Interspike Interval Reading Unit (IRU)}

Using these time delay units we now construct a circuit that can
be trained to be selective for a chosen ISI sequence by repeated
presentation of given ISI sequence. The main idea is that an ISI
sequence starting at time $t_{0}$ and with spikes at times:
$t_{0},t_{0}+T_{0},t_{0}+T_{0}+T_{1},\cdots,t_{0}+\sum_{j=0}^{N-1}T_{j}$
induces a replica of itself using time delay units with time
delays $\tau_{0},\tau_{1},\cdots,\tau_{N-1}$. This replica
sequence is compared to the original sequence and, if
$|\sum_{j=0}^{i-1} (T_{j}-\tau_{j})|>\delta$ ms, where $\delta$ is
the resolution of spike detection, then an inhibitory synaptic
plasticity rule, as we discuss below, modulates delay unit $i$, so
that this difference is reduced towards the tolerance limit of
$\delta$ ms. When $|\sum_{j=0}^{i-1} (T_{j})-\tau_{j})|\le
\delta\,$ ms for each ISI in the training sequence, the IRU is
considered trained. We choose $\delta=1$ ms, which reflects the
width of actual action potentials; mathematically we could choose
any $\delta >0$ as our convergence criterion (Appendix E).
Training of the delay units in IRU is clearly a function of the
learning rule implemented to adjust the $\tau_{i}$'s.

The complete circuit for an IRU is shown in Figure ~\ref{IRU}. The
selected ISI sequence enters the ``gateway" unit that passes on
only the first spike. This effectively synchronizes the clocks of
the first delay unit. The main purpose of this synchronization of
the delay units is to prevent the other spikes in the input ISI
sequence from interfering with the operation of these time delay
units. The gateway unit is effectively closed after the first
spike is received, and it is reset by an inhibitory signal sent by
the last time delay unit when the replica ISI sequence has been
created and passed on to the ``detection" unit. The original ISI
sequence is also passed directly to the time delay units for
training and to the detection unit for comparison with the
replica.

The gateway unit is shown in Figure ~\ref{gateway}. It is
comprised of an excitatory neuron which receives the input ISI
sequence and passes a first spike, at time $t_{0}$, on to the
first time delay unit. It also excites a bistable inhibitory
neuron or small neural circuit  which moves into an oscillatory
state and effectively turns the excitatory neuron off until it, in
turn, is set back into its rest state by a signal from the last
time delay unit.

The detection unit is comprised of a neuron (or small circuits of
neurons) that fires when two spikes arrive within 1 ms of each
other.  It remains below its action potential threshold otherwise.
There are many ways to accomplish this, and in the appendix we
give an example of the detection unit we used in our construction
of IRU.

Finally we need  a mechanism to train each of the time delay units
to adjust $|\sum_{j=0}^{i-1} (T_{j}-\tau_{j})|\le 1\,$  ms.
Equivalently we require a mechanism to adjust each delay unit such
that $|T_{i-1}-\tau_{i-1}| =|\delta_{i-1}| \le 1$ ms, for each of
$i^{th}$ delay unit such that $|\sum_{j=0}^{i-1}\delta_{j}|\le 1$
ms. This is done by presenting the target ISI sequence to the
C-neuron of each time delay unit. The C-neuron of the i$^{th}$
time delay unit fires an action potential at time
$t_{0}+\sum_{j=0}^{i-1}\tau_{j}$ and receives a spike from the
input ISI sequence at $t_{0}+\sum_{j=0}^{i-1}T_{j}$. If these are
more than 1 ms apart,  the inhibitory spike timing dependent
plasticity rule will adjust it to be less than or equal to 1 ms as
discussed in next section. There we discuss the experimentally
observed spike timing dependent plasticity rule for an inhibitory
synapse and explain how it plays a role in training the IRU to
accurately detect the input ISI sequence.

\section{IRU learning}

A spike timing dependent plasticity rule for inhibitory synapses
has recently been observed in layer II of entorhinal cortex by
Haas, {\em et al}~\cite{Haas}. It gives the change in inhibitory
synaptic conductance associated with the arrival of a spike at
$t_{pre}$ at the presynaptic terminal and spiking of a
postsynaptic neuron at $t_{post}$. As a function of $\Delta t =
t_{post} - t_{pre}$, the change in synaptic conductance ${\Delta
g_{I}(\Delta t)}$ normalized by the baseline synaptic conductance
$g_{I0}$ is given by \be
 \frac{\Delta g_{I}(\Delta t)}{g_{I0}}=
\alpha^{\beta} \Delta t |\Delta t|^{\beta-1}\exp(-\alpha|\Delta
t|). \ee An empirical fit to the data gives $\beta \alpha \approx
5$ to $10$ ms$^{-1}$. We have chosen $\beta = 5$ and $\alpha=1\,$
ms $^{-1}$ in the computations we report here (See the graphic
inset in figure ~\ref{learnrule}). This empirical learning rule
allows tuning of the inhibitory synapse from B$\rightarrow$ C in
the delay unit of the IRU. We identify $\Delta t$ as the time
between the action potential generated at neuron C corresponding
to $t_{pre}$, the time of C spike induced by presynaptic
stimulation of neuron B and $t_{post}$, the time of receipt of the
appropriate spike at neuron C, in the selected ISI sequence. For
the first time delay unit, for example, this is $T_0 -
\tau_0(R_0)$ (figure ~\ref{IRU}).

Consider the i$^{th}$ delay unit and suppose the previous i-1
units have been adjusted such that $\left
|\sum_{j=0}^{i-2}(T_{j}-\tau_{j})\right |\le \delta$, where
$\delta$ is the resolution scale for ISI detection. We now need to
adjust the time difference $|\Delta t|=|T_{i-1}-\tau_{i-1}|$
between the ISI sequence time $t_{0}+\sum_{k=0}^{k=i-1}T_{k}$ and
the present value of the time delay unit output at
$t_{0}+\sum_{k=0}^{k=i-1}\tau_{k}$  for the i$^{th}$ unit. Before
any adjustment this $|\Delta t|$ is greater than $\delta$. As an
example this is shown in Figure ~\ref{learnrule}, for a scenario
in which we start with the i$^{th}$ delay unit initially tuned to
produce a delay of $\approx$ 56.5 ms, and the target ISI it needs
to detect is around 47.5 ms resulting in $|\Delta t|=9$ ms which
is greater than $\delta$. $\delta$ $\approx 1\,$ ms.

The learning rule, shown in the inset of Figure~\ref{learnrule},
requires $g_{Ii}$, to be modified until $|\Delta t|<\delta$. If N
represents the trial number of the repeated presentations of our
desired ISI sequence, then we have a one dimensional map for the
evolution of $g_{Ii}$, given by $g_{Ii}(N+1)=f(g_{Ii}(N))$, where
$f(x)=x+\Delta x$. We are looking for a solution
$g_{Ii}=g^{*}+\delta_{R}$, of the map corresponding to the
resolution window $\delta$ for ISI detection. The case shown in
Figure~\ref{learnrule} corresponds to the situation when $\Delta t
<-\delta$. The following analysis can easily be extended to the
case when $\Delta t >\delta$.

The inhibitory synapse learning rule, for scenario shown in Figure
~\ref{learnrule} implies
\begin{eqnarray*}
\Delta g&=&\alpha^{\beta}\Delta t|\Delta t|^{\beta-1}\exp(-\alpha|\Delta t|) \nonumber \\
&=& - (\alpha |\Delta t|)^{\beta} \exp((-\alpha |\Delta t|)) , \left(\Delta t <-\delta <0\right) \nonumber \\
&<&0
\end{eqnarray*}
where $\Delta t=T_{i}-\tau_{i}(g)$. Since $\Delta g <0$ implies
the value of g over the next iteration decreases, and since
$\frac{d\tau (g)}{dg} >0$ (Figure ~\ref{delay}) implies the $\tau$
value decreases, this results in $\Delta t$ approaching zero from
the right. As a result each repeated presentation of the desired
ISI sequence will drive $g_{Ii}$  towards $g^{*}$.

It can be shown that for any learning rule $\Delta g(\Delta t)$,
with $\tau(g)$, such that $\frac{d\tau}{dg}>0$ satisfying, the
condition, $\Delta g >0\hspace{.2cm} \forall \hspace{.2cm} \Delta
t >0$ and $\Delta g <0 \hspace{.2cm}\forall  \hspace{.2cm}  \Delta
t <0$, the convergence to desired ISI sequence will occur in
finite number of steps (See appendix E ).

For the particular case of the learning curve used here,
abstracted from empirical fit to observed experimental data on
inhibitory synapses, the number of steps for convergence will
depend on the criteria $\delta$ chosen for learning to stop. We
give here several examples of learning for $\delta$ chosen to be 1
ms. In Figure ~\ref{Converge}, in particular, we show how the rate
of convergence for learning changes as function of $\delta$, for
IRU training on the ISI sequence $T_{i}$ to be 46.0 , 51.7 , 56.3
and 61.7 ms, respectively with the initial delays $\tau_{i}(N=0)$
for each of the delay units in IRU set at 52.68 ms, which is done
by setting initial R value for each unit.

\par In Figure ~\ref{train1}  we give an example
of the training of an IRU with four time-delay units by a selected
ISI sequence. The input ISI sequence is fed into the gateway unit
which initializes activity in the delay unit. This ISI sequence
also enters neuron C of each delay unit. In neuron C the ISI
sequence participates in modulating the inhibitory synaptic
strength of the B$\rightarrow$C synapse. At the same time the ISI
sequence is also fed into the detection unit which detects the
presence or absence of two spikes within our chosen resolution of
1 ms, thereby signaling detection of an input spike when there is
a coincidence. In Figure ~\ref{train1} the ISIs in the training
sequence, the $T_i$, are taken to be 50, 60, 55, and 58 ms
respectively. The initial time delays $\tau_i(N=0)$ in the four
time delay units are chosen to be 61.51, 57.55, 55.41 and 58.63
ms.

Since the IRU has time delay units connected in a chain, any
modification in delay unit 1, for example, reflects itself in
delay unit 2, and on down the chain. It is helpful to understand
this connection among the delay units by expressing the way any
given unit achieves its proper delay subtracting off the activity
in earlier delay units. Figure ~\ref{train1} shows the time delays
as a function of the number of presentations of the selected ISI
sequence in three different forms. Figure ~\ref{train1}a displays
the error in all delay units (i=1 to 4) \bea {\mathbf
E^{IRU}_{i-1}}(N) &=& \sqrt{((t_{0}+\sum_{j=0}^{i-1}\tau_{j}(N))
-(t_{0}+\sum_{j=0}^{i-1}T_{j}))^2 } \nonumber \\
&=& \sqrt{(\sum_{j=0}^{i-1}(\tau_j(N) - T_j))^2}. \eea as a
function of the training sequence presentation number N. When the
training is perfectly completed so $\left|\tau_{j}(N) - T_{j}
\right|\approx 0$, for each of the delay unit this error goes to
zero. The actual training continues until the error in each of the
four delay units is within 1 ms of the threshold for spike
detection by the detection unit.

In Figure ~\ref{train1}b we plot the actual delay produced by the
$i^{th}$ delay unit as it receives input spikes from  the previous
delay unit, (i-1)$^{st}$, at time $t_{0}+\sum_{j=0}^{i-2}\tau_{j}$
and produces spike output at time
$t_{0}+\sum_{j=0}^{i-1}\tau_{j}$. In Figure ~\ref{train1}c, we
plot the difference \bea {\mathbf \Gamma^{IRU}_{i-1}}(N) &=&
(t_{0}+\sum_{j=0}^{i-1}\tau_{j}(N))-
(t_{0}+\sum_{j=0}^{i-2}T_{j}) \nonumber \\
&=& \tau_{i-1}(N) + \sum_{j=0}^{i-2}(\tau_j(N) - T_j) \eea
corresponding to the evolution of the $i^{th}$ ISI in the IRU as a
function of the training sequence presentation number. This
quantity tells us how the value of the (i)$^{th}$ time delay unit
depends on the adjustments still taking place in the units earlier
in the chain of delay units. When all the ISI's have been
detected, this number for each delay unit will be within  $\delta$
ms of the actual ISI, as can be seen in Figure ~\ref{train1}c. In
the subsequent training examples, we only plot ${\mathbf
\Gamma^{IRU}_{i-1}}(N)$, as shown in Figure ~\ref{train1}c, and
the corresponding error, ${\mathbf E^{IRU}_{i-1}}(N)$, as shown in
Figure ~\ref{train1}a.

The rate of change of the inhibitory synaptic conductance is
determined by a scale  factor of $g_{I0}$, set to
$0.1e^{\beta}/\beta^{\beta} $ here, in the expression for $\Delta
g_{I}(\Delta t)$. With this choice it takes about 88 presentations
of the training ISI sequence to reach the desired accuracy for all
time delays. In the simple training rule we use, changing the
dimensionless number $g_{I0}$ from the value chosen above, to a
large number increases the convergence rate as it simply scales
time.

In Figure ~\ref{train2} we give an additional example of training
an IRU in a noise free environment. We show in Figure
~\ref{train2} the results of training an IRU when all time delays
are initially set to 52.13 ms, and the input ISI sequence has
ISIs:  46, 51.7, 56.3, and 61.7 ms. Figure ~\ref{train2}a shows
the evolution of ISI from each delay unit as a function of the
number of presentations of the selected ISI training set. In
Figure ~\ref{train2}b we show the detection unit time series in
the untrained case, We see three subthreshold responses of the
detection unit neurons associated with the input spikes, and we
see one action potential associated with one of the untrained time
delays having by chance been set within 1 ms of an input ISI.
Figure ~\ref{train2}c displays the detection unit output in a
partially trained IRU when three ISIs in the time delay units lie
within 1 ms of ISIs in the input ISI sequence. Finally in Figure
~\ref{train2}d we show the fully trained IRU where we see four
detection unit action potentials associated with the
``coincidence," within 1 ms, of the time delays and the ISIs in
the time delay units.

\section{Performance of an IRU in a noisy environment}

We have shown that an IRU is able to adjust its time delay units
in a manner allowing detection of a specific noise free input ISI
sequence. This is a ``clean" environment allowing us to examine
properties of IRU performance in absence of noise which is always
present in any realistic biological setting. A thorough,
systematic examination of IRU performance in a realistic noisy
setting is the subject of another investigation. Here we confine
ourselves to two instances of noise in the IRU detection.

In the first instance, we allow the training ISI sequence to have
jitter on each of the ISIs in the sequence. This is the case where
an input sensory signal which would ideally be represented by a fixed set of
ISIs is now subject to noise in the neural circuitry or other environmental perturbations.
The firing times of neurons in the processing circuitry transforming analog environmental
signals to the ISI sequences would certainly contribute to this form of noise.

Figure ~\ref{Noise1}a shows the training of the time delay units
in an IRU in a noise free environment. In the top panel of Figure
~\ref{Noise1} we present the error in ISI detection ${\mathbf
E^{IRU}_{i-1}}(N)$ at each delay unit in an IRU as a function of
training number. In the bottom panel of Figure ~\ref{Noise1}a we
show the actual evolution of the ISI's for each delay unit as a
function of training number. The initial time delay for each unit
is chosen to be $\tau_{i}(0)=52.68$ ms, and the ISI training
sequence consists of the ISIs: $T_{1}=46.0$, $T_{2}=51.7$,
$T_{3}=56.3$, and $T_{4}=61.7$ ms.

In Figure ~\ref{Noise1}b we have introduced a jitter in each spike
of the input ISI sequence uniformly distributing the spike time
over $\pm 2$ ms around the mean values 46.0, 51.7, 56.3, and 61.7
ms. The IRUs train each delay unit until the time delay is within
1 ms of the input ISI. The training stops once the detection unit
has produced 4 spikes corresponding to detection of each ISI.
Since that might happen within $\pm$ 2 ms jitter of each spike,
each of the output time delays in the bottom panel of Figure
~\ref{Noise1}b is only reliable within the 2 ms jitter.

The process of repeatedly presenting the ISI sequence of interest
to the IRU and altering the B$\to$C inhibition each iteration is
dynamically a discrete time map of the IRU plus our $\Delta g_I$
rule. We are seeing in this example a representation of the
stability of a fixed point of that iterated map system. This will
appear again in our upcoming examples. We have not studied the
basin of attraction of this fixed point or other long time
dynamical behaviors of this map.

As a second instance of a noisy training environment, we have also
investigated the situation in which an ``extra" spike randomly
appears in our deterministic ISI sequence. In Figures
~\ref{Noise2}a and ~\ref{Noise2}b we show two cases of learning
the same underlying deterministic ISI sequence as in Figure
~\ref{Noise1} but now with an additional spurious spike present in
the 3rd ISI interval with probability p=0.25 or p=0.5. We see that
IRU is still able to learn the correct ISI sequence in about 101
presentations of the sequence with a random extra spike.

In terms of the possible existence of IRU in birdsong system, when
the bird is deafened, the precise timing of the spikes resulting
from sensory input is lost as it no longer receives auditory
input, and as a result the song degenerates. In the context of an
IRU this means that the IRU trained on the tutor song and
maintained on the bird's own song no longer receives the specific
ISI sequence representing that song. Instead, in a manner we do
not know precisely, it will receive random sequences of spikes
with ISIs having little or nothing to do with birdsong but
representing neural environmental noise.

In Figure ~\ref{Noise3} we show the training of an IRU for 20
presentations of the desired ISI sequence with ISIs of 50, 60, 55,
and 58 ms. This is followed by 60 further presentations of a spike
sequence with the same mean ISIs but perturbed by random normally
distributed noise with zero mean and RMS variation of 5 ms. This
is then followed by 60 presentations of the noise free initial ISI
sequence. We see how the IRU is initially effectively training
itself on the deterministic sequence, then it loses this
connection with the desired ISIs when random spikes are received.
Finally it recovers to train properly when the noise is removed.

\section{Discussion}
We have discussed a circuit composed of biologically motivated
neurons and biologically motivated synaptic connections designed
to respond selectively to a particular ISI sequence. We begin with
the possibility of constructing a time delay circuit with the
fewest possible number of neurons, two in this case, and then
describe a three neuron time delay circuit model, abstracted from
the birdsong system. The untrained IRU network has a set of time
delay units constructed from three neurons each of which has an
adjustable time delay set by the strength of an inhibitory
synapse. These time delays are themselves set by a synaptic
plasticity rule which compares the ISIs in the ISI sequence of
interest to the time delays in the time delay units and adjusts
the latter until they match the presented ISIs within a certain
error, taken here to be 1 ms.

The construction of the overall circuit (an IRU) to read ISIs in
the chosen sequence is quite general and is not solely connected
with the observations which motivated its construction. It could
be, though we do not have anatomical or electrophysiological
evidence for this at this time, that such circuitry could be used
generally to recognize the specific ISI sequences produced by
sensory systems in response of environmental stimuli.

It seems clear that some circuit of this kind, whether or not it
is the one we construct here, may well be utilized by animals for
recognition of important sensory inputs. Those inputs are
transformed by the sensory system into spike sequences, and in
situation when all the spikes produced are identical, all the
information is represented in the spike sequence. Reading those
sequences in pre-motor or decision processing is required for
various functional activities.

Our time delay circuit is constructed by analogy with one found in
the AFP of the song system of songbirds. It is slightly simplified
by the elimination of an output nucleus that is used for detailed
tuning of the pre-motor responses in RA~\cite{Kao}, and it is
represented by a three neuron circuit. Its instantiation in a
biological system could, of course, use many more neurons as in
the case of birdsong.

The time delay of this three neuron circuit is set in overall
scale by the membrane capacitance of the neurons, and it is tuned
in detail by the strength of one of the internal inhibitory
connections. This strength is set by inhibitory synaptic
plasticity using rules recently observed by~\cite{Haas}. The
regime of operation of the IRU utilizing such time delay circuits,
is governed by $\tau(R)$, and the IRU can train itself on ISIs in
the interval $\left[\tau(R_{L}), \tau(R_{U})\right]$

We gave examples of the training of IRU network using an
inhibitory synaptic plasticity rule. The outputs of the untrained
network, a partially trained network, and a fully trained network
which can recognize four ISIs are shown in Figures
~\ref{train2}b-d. A ``detection unit" which fires an action
potential when a spike from a neuron C of the time delay unit
arrives within 1 ms of a spike from the selected input ISI
sequence produces no spikes for the untrained unit and reliably
produces four spikes associated with the correct ISIs after
training using synaptic plasticity. No spike in the detection unit
is produced by the initial spike in the input ISI sequence; the
detection unit responds only to correct ISIs after training using
synaptic plasticity.

Our analysis does not address the response of an IRU to a desired
ISI sequence when it is embedded in an environments with many
extraneous spikes, and it does not address the reliability of the
synaptic connections as a potential source of error in reading ISI
sequences. For example, suppose the second spike in a sequence is
missing because of failure of a synapse to fire when expected. How
does the IRU performance degrade under such circumstances. It may
be that the actual biological environments in which IRUs operate,
assuming them to be present, require a statistical measure of
detection efficacy. This would need to be carefully connected with
the dramatic response of HVC neurons projecting to RA when
stimulated by BOS as seen in the work of Coleman and
Mooney~\cite{Coleman1}. In connecting the ideas here about reading
ISI sequences with models of the NIf and HVC nuclei in the avian
birdsong system one will be able to address this and other issues
in a quantitative fashion.

\clearpage

\appendix

\section{Time Delay Unit of the IRU}
In our three neuron model of the time-delay circuit we used HH
conductance based models. In these equation there appear the
sodium and potassium ion channels activation variables m(t) and
n(t) and the sodium channels inactivation variable h(t). These
satisfy, with $X(t)=\{m(t),h(t),n(t)\}$
\begin{eqnarray*}
\frac{dX(t)}{dt}&=&\alpha_{X}(V(t))(1-X(t))-\beta_{X}(V(t))X(t)
\end{eqnarray*}
where V(t) here refers to the membrane voltage of the neuron, A, B, or C. We use, with V given in mV:
\begin{eqnarray*}
\alpha_{m}(V)&=&\frac{.32(13-(V-V_{th}))}{(\exp((13-(V-V_{th}))/4.)-1)} \nonumber \\
\beta_{m}(V)&=&\frac{.28((V-V_{th})-40)}{(\exp(((V-V_{th})-40)/5)-1)} \nonumber \\
\alpha_{h}(V)&=&.128\exp((17-(V-V_{th}))/18) \nonumber \\
\beta_{h}(V)&=&4.0/(1+\exp((40-(V-V_{th}))/5)) \nonumber \\
\alpha_{n}(V)&=&\frac{0.032(15-(V-V_{th}))}{(\exp((15-(V-V_{th}))/5)-1)}\nonumber \\
\beta_{n}(V)&=&.5\exp((10-(V-V_{th}))/40)
\end{eqnarray*}
and $V_{th}$=-65mV.

The various parameters chosen for the circuit are : The membrane
capacitance is $C_M = 1.0 \frac{\mu F}{cm^2}$ . The maximal
conductances,of the ionic currents in units of $mS/cm^{2}$ are,
$g_{Na} = 215, g_{K}=43, g_{L}=.813$. The reversal potentials in
units of mV are $V_{Na} = 50, V_{K} = -95,$ and $V_{L}=-64$. The
excitatory synaptic conductances are $g_{EA}=1.0 mS/cm^{2}$,
$g_{EB}=1.0 mS/cm^{2}$. The inhibitory synaptic conductance,
$g_{I}=1 mS/cm^{2}$. The scaling factor $R^{0}=50.0$ and $R$,
varies as given in text. $V_{revE}=0\, mV$ , and $V_{revI}=-80\,
mV$. $\tau_{E}=1.0\, ms$, $S_{1E}=1.5$, $\tau_{I}=1.2\, ms$,
$S_{1I}=4.6$. The DC currents in the neurons are taken as
$I^{DC}_{A}=0.0 \mu A/cm^{2}$, $I^{DC}_{B}=1.97 \mu A/cm^{2}$ and
$I^{DC}_{C}=1.96 \mu A/cm^{2}$

\section{The Gateway Unit}

We model the gateway unit, as a bistable neuron coupled to
standard HH model described above. The neuron A, receives the
input ISI sequence beginning at time $t_{0}$. A spike from neuron
A triggers the bistable neuron B, which is driven into a stable
periodic spiking regime. Once the bistable neuron starts to fire,
it, in turn, inhibits, neuron A and prevents it from responding
any further to input ISI spikes. The membrane voltage of neuron A
satisfies the following dynamical equation,
\begin{eqnarray*}
C_{M}\frac{dV_{A}(t)}{dt}&=&g_{Na}m(t)^{3}h(t)(V_{Na}-V_{A}(t)) \nonumber \\
&+&g_{K}n(t)^{4}(V_{K}-V_{A}(t))+g_{L}(V_{L}-V_{A}(t)) \nonumber \\
&+&g_{I0}S_{I}(t)(V_{revI}-V_{A}(t)) \nonumber \\
&+&g_{E0}S_{E}(t)(V_{revE}-V_{A}(t))+I_{DC0}
\end{eqnarray*}
where the symbols repeated from the description of the time delay
unit have same meaning and values. In addition, the strength of
inhibitory synapse from bistable neuron B onto A is $g_{I0}= 100 $
$mS/cm^{2}$ and the strength of excitatory synaptic input to
neuron A from ISI sequence source $g_{E0}=1$ $mS/cm^{2}$.
$I_{DC0}=1.0 \, \frac{mA}{cm^2}$, such that neuron is sitting at
resting potential of -63 mV.

The dynamics of neuron B which is in a bistable regime is given by

\begin{eqnarray*}
C_{M}\frac{dV_{B}(t)}{dt}&=&g_{Na}m_{\infty}(t)(V_{Na}-V_{B}(t)) \nonumber \\
&+&g_{K}n(t)(V_{K}-V_{B}(t))+g_{L}(V_{L}-V_{B}(t)) \nonumber \\
&+&I_{syn}(t)+I_{DC1}
\end{eqnarray*}
where the gating variables $X(t) = \{n(t),m(t)\}$ satisfies the following kinetic equation,
\begin{eqnarray*}
\frac{dX(t)}{dt}&=&\frac{X_{\infty}(V(t))-X(t)}{\tau_{X}}
\end{eqnarray*}
and  the activation function, $X_{\infty}(V)=\{m_{\infty}(V),n_{\infty}(V)\}$, has the
following functional dependence on voltage V,
$$X_{\infty}(V)=1/(1+\exp((V_{X}-V)/k_{X}))$$

$I_{syn}(t)$ is the excitatory synaptic current from neuron A on
to neuron B, which is taken to be,
$I_{syn}(t)=g_{E1}S_{E}(t)(V_{revE}-V_{B}(t))$, which represents
AMPA type excitatory synaptic current.  Finally $S_{E}(t)$ is
taken to satisfy the following first order kinetic equation,
\begin{eqnarray*}
\frac{dS_{E}(t)}{dt}&=&\frac{S_{0}(V_{pre}(t))-S_{E}(t)}{\tau_{E}(S_{1E}-S_{0}(V_{pre}(t)))}.
\end{eqnarray*}
where as described in the main text we have $\tau_{E}=1$ ms and
$S_{1E}=0.5$, giving the docking time of .5 ms and undocking time
of 1.5 ms. Finally, the various parameters appearing in these
equations are $C_{M}=1\frac{\mu F}{cm^2}$; $g_{Na} = 20, g_{K}=10,
g_{L}=8.0$, in units of $mS/cm^{2}$; $V_{Na} = 60, V_{K} = -90,$
and $V_{L}=-80$, in units of mV; $V_{m}=-20$mV, $k_{m}=15$,
$V_{n}=-25$mV, $k_{n}=5$, $\tau_{n}=1\,$ms. The conductance of the
excitatory synaptic connection is $g_{E1}=1\, mS/cm^{2}$.
$I_{DC1}=4\, mA/cm^{2}$, such that the neuron B is at resting
potential of around -62 mV.

\section{One and Two neuron Delay Circuit}

All the parameters for one neuron model described in the main
article are same as the ones used for construction of neural units
for three neuron model as described in appendix A. For the two
neuron model the bistable neuron B, satisfies the same set of
dynamical equations as presented above for the bistable neuron B
of the gateway unit. Similarly the type I neuron of the two neuron
delay model shares same set of equations as described for the
neuron A of the gateway unit. All the dynamical equations for the
type 1 neuron and the bistable neuron described in the gateway
section are same for description of the two neuron model, except
for the excitatory synapse, which is modelled as NMDA type
excitatory synapse, necessitated by the slow decay times, required
for neuron B to provided enough excitatory input to neuron A for
it to spike. The excitatory synaptic current from $B\rightarrow
A$, is given by $I(t)=g_{E}S_{E}(t)B(V(t))(V_{revE}-V(t))$, where
$B(V(t))=1.0/(1+0.288[Mg^{2+}]\exp(-0.062V(t)))$ and
$[Mg^{2+}]=1$mM. The strength of inhibitory connection from neuron
A to neuron B is 1 mS/cm$^{2}$ and the strength of excitatory
synaptic input from source onto neuron B at time $t_{0}$ is 0.1
mS/cm$^{2}$.
\section{Detection Unit}

The detection unit model used in our IRU, is a type I neuron model
as discussed above for the delay unit in appendix A. It is tuned
to respond to two input spikes arriving within a millisecond of
each other. In Figure ~\ref{Detect}b top panel, we show the
response of the detection unit when it receives two input spikes,
2 ms apart. The total integrated input at given time delay between
these two inputs is not sufficient to push the neuron beyond its
spiking threshold and the neuron responds only with an EPSP.
However, when two input spikes, are sufficiently close in time,
here within 1ms, as shown in Figure ~\ref{Detect}b bottom panel,
the total integrated input is sufficient to push the neuron above
its spiking threshold. Now it responds with a spike after a time
delay of around 10 ms, governed by the integration time constant
of the neuron dynamics.

\section{Linear Map for Learning Rule}

In this section we consider an approximation of the learning rule
for evolution of inhibitory synapse, and the delay produced by
each delay unit as function of strength of inhibitory connection
from $B\rightarrow C$, as shown in Figure ~\ref{LinearRule}a and
~\ref{LinearRule}b, and compute an analytical expression for
number of training sequence steps required for delay unit to
detect a spike within $\delta$ ms resolution of actual ISI. The
approximation shown in Figure 16a and ~\ref{LinearRule}b results
in the following,
\begin{eqnarray*}
\frac{\Delta g_{I}}{g_{I_{0}}}&=&b\Delta t, \hspace{.2cm} ( |\Delta t| \le A )\nonumber \\
&=& -b\Delta t+2Ab,  \hspace{.2cm} (A < |\Delta t| \le 2A ) \nonumber \\
&=&0,\hspace{.2cm} (|\Delta t| >2A) \nonumber \\
\end{eqnarray*}
with, $\tau=a g_{I}+c \hspace{.2cm} ( g_{L}\le g_{I}\le g_{U} )$
and $\tau=0\hspace{.2cm} (g_{I}<g_{L}) $ and $\tau=\tau(g_{U})
\hspace{.2cm}(g_{I}>g_{U}) $. The fact that $\Delta g=0$ for
$|\Delta t|>2A$ implies that, learning will occur only for $\Delta
t$ in the range of $\pm$ 2A ms, or in the map $f(g_{I}(N))$ the
allowed variation in $g_{I}(N)$ is from (T-c-2A)/a to (T-c+2A)/a.
Depending on the initial inhibitory synaptic strength of
$g_{I}(0)=g_{0}$, with above linear learning rule, the number of
steps for the delay unit to set its delay output within $\delta$
ms of actual ISI time that it needs to detect, $N(\delta)$, can be
obtained as follows,

The trivial case of initial condition being within the $\delta$
window of actual ISI, results in $N(\delta)=0$. In the situation
when $|\Delta t| \le A$, the number of training iterations
required for learning, is given by $N(\delta)=1+n_{1}$, where
$n_{1}$, can be computed as follows. We begin in the region
$\Delta t=T-\tau_{0} \le A$, which corresponds to initial
inhibitory synaptic strength lying in the interval,$(T-c-A)/a \le
g_{I}(0)\le(T-c-\delta)/a$. At each iteration step i, as shown in
the example path in Figure 16c, $g(i)$, increments by amount,
$\Delta g_{I}(i)/(1-ab^{'})$, where $b^{'}=bg_{I_{0}}$. The total
number of integer steps required for $g_{I}(0)=g_{0}$, to evolve
to within $\delta$ ms window of T, is then given by
\begin{eqnarray*}
n_{1}&=&\left[\frac{\log\left(\frac{\delta}{(|T-\tau_{0}|)}\right )}{\log(1-ab^{'})}\right]
\end{eqnarray*}

where $\tau_{0}=ag_{0}+c$ and $[x]$, is the largest integer less than or equal to $x\in \mathbb{R}$.

It is important to note the factor of $b^{'}$ appearing in the
denominator of above equation. In the scheme of learning rule we
have used in the main calculations, as $\Delta t\rightarrow 0$,
$b^{'}$, which represents the slope of the learning curve

approaches 0, and as we can see from above, theoretically the
exact convergence of learning, i.e., $\Delta t=0$ requires,
infinity of training steps.

In situation when the initial condition is such that $A<|\Delta
t|<2A$, the number of training iterations required is given by
$N(\delta)=2+n_{1}+n_{2}$, where
\begin{eqnarray*}
n_{1}&=&\left[\frac{\log\left(\frac{A}{|2A+\tau_{0}-T|}\right )}{\log(1+ab^{'})}\right] \\
n_{2}&=&\left[\frac{\log\left(\frac{\delta}{|(T-c)-a\widetilde{g}|}\right )}{\log(1-ab^{'})}\right] \\
\widetilde{g}&=&g_{0}{(1+ab^{'})}^{n_{1}+1}+\frac{\left((1+ab^{'})^{n_{1}+1}-1)(2A-(T-c)\right)}{a} \\
\end{eqnarray*}

Linear fit to the learning rule used in the main text, which is
abstracted from empirical fit to entorhinal cortex data, gives,
$A=5 $ms, $b^{'}=0.02 $ms$^{-1}$ and linear $\tau(g)$ curve
implies, $a=0.9$ ms $c=42.58$ ms, $g_{L}=.75$ and $g_{U}=22.0$. In
this approximation the maximum number of steps will correspond to
beginning with $\Delta t$ error of $\pm 2A=10 ms$. For a
particular case of T=60 ms, and beginning with $\tau_{0}=51 ms$,
giving $\Delta t=9 ms$, and taking $\delta=1ms$, solving the above
equation gives, N=179, as total number of training cycles for the
delay unit to learn.

\section*{Acknowledgments}
This work was partially funded
by the U.S. Department of Energy, Office of
Basic Energy Sciences, Division of Engineering and Geosciences,
under Grants No. DE-FG03-90ER14138 and No. DE-FG03-96ER14592; by a
grant from the National Science Foundation, NSF PHY0097134,
and by a grant from the National Institutes of Health, NIH R01 NS40110-01A2.
HDIA and SST are partially supported by the NSF sponsored Center for Theoretical Biological
Physics at UCSD.
We are very appreciative of comments by Leif Gibb, Marc Schmidt, Jon Driscoll, and Dan Margoliash
on early drafts of this paper. Their observations and suggestions helped us improve
the paper significantly.

\Large{\bf {Figures}}

\begin{figure*}[ht!]
\begin{center}

\includegraphics[width=7in,scale=1,angle=0]{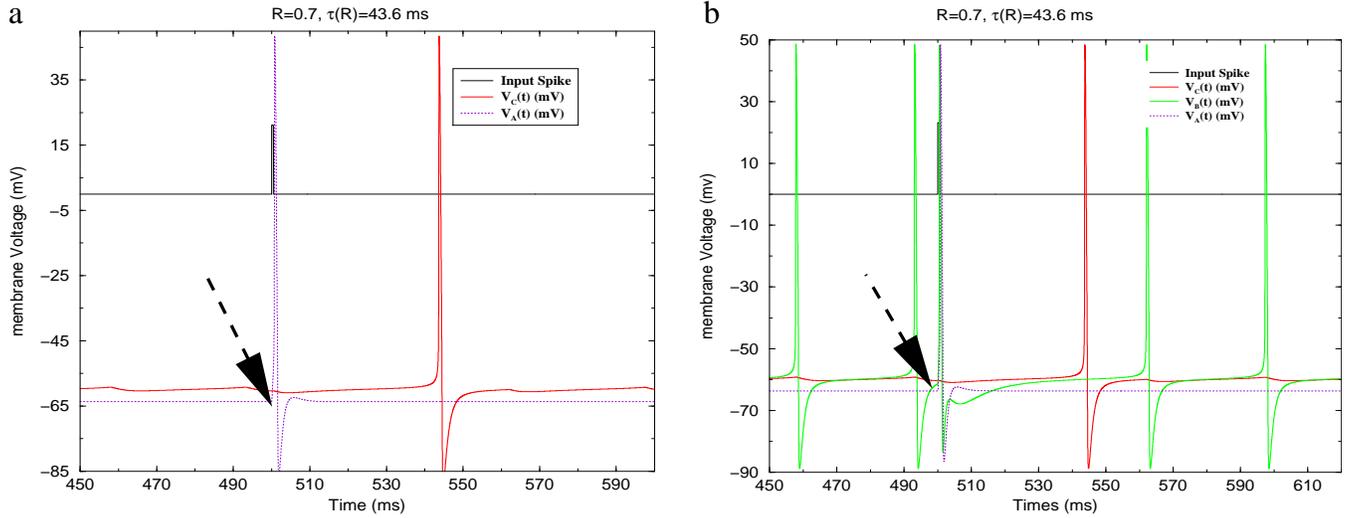}
\caption{(a) For R = 0.7 we show the membrane voltages of neuron A
(blue) and neuron C (red) in response to single spike input
(black) arriving at neuron A and neuron B at time $t_{0}=500\,$
ms. For $R = 0.7$ we see the output spike from neuron C occurring
at $t=543.68\,$ ms, corresponding to $\tau (R)=43.6\,$ms. (b) For
R =0.7 we again show the membrane voltages of neuron A (blue) and
neuron C (red), and in addition now display the membrane voltage
of neuron B (green). A single spike input (black) arrives at time
$t=500$ ms. We see that the periodic action potential generation
by neuron B is reset by the incoming signal. In Figures 2a and 2b
the arrow indicates the time of the spike input to units A and B
of our delay unit. \label{membrane1}}
\end{center}

\end{figure*}

\begin{figure*}[ht!]
\begin{center}

\includegraphics[width=5in,scale=1,angle=0]{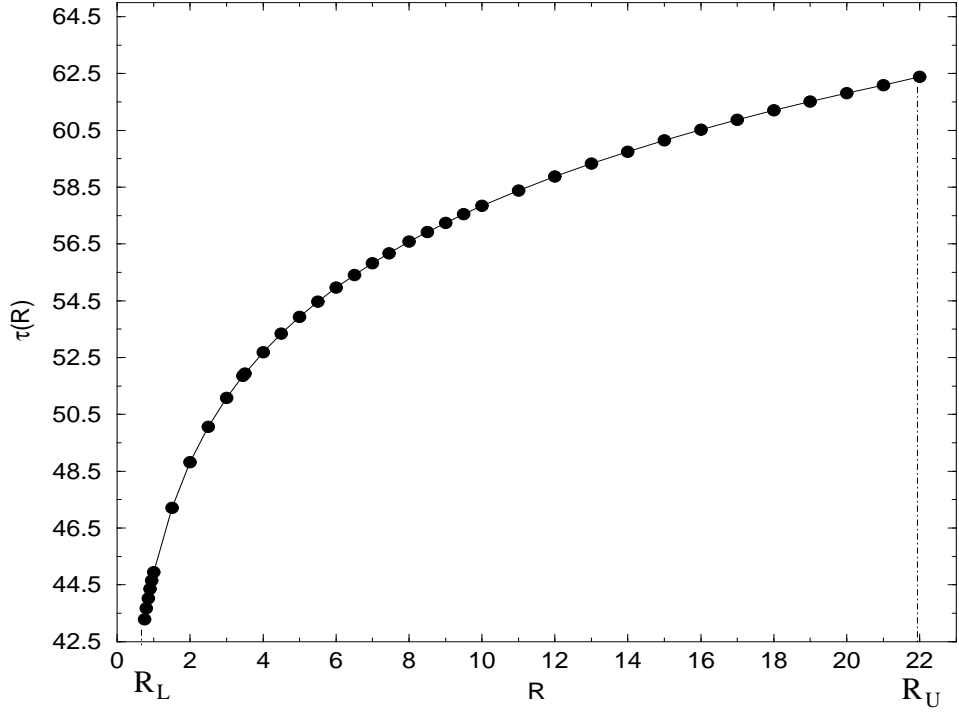}
\caption{The delay $\tau(R)$ produced by the time delay unit as a
function of $R$, the strength of the inhibitory synaptic
connection $B\rightarrow C$. All other parameters of the time
delay circuit are fixed to values given in the appendix. For
$R<R_{L}$ the inhibition is too weak to control the precise
spiking of neuron C, locked in with respect to signal input spike
to the delay unit. For $R>R_{U}$ the inhibitory synapse is so
strong that the neuron does not produce any action potential,
effectively the delay out of the network is infinity.
\label{delay}}
\end{center}
\end{figure*}

\begin{figure*}[ht!]
\begin{center}
\includegraphics[width=4.65in,scale=1,angle=0]{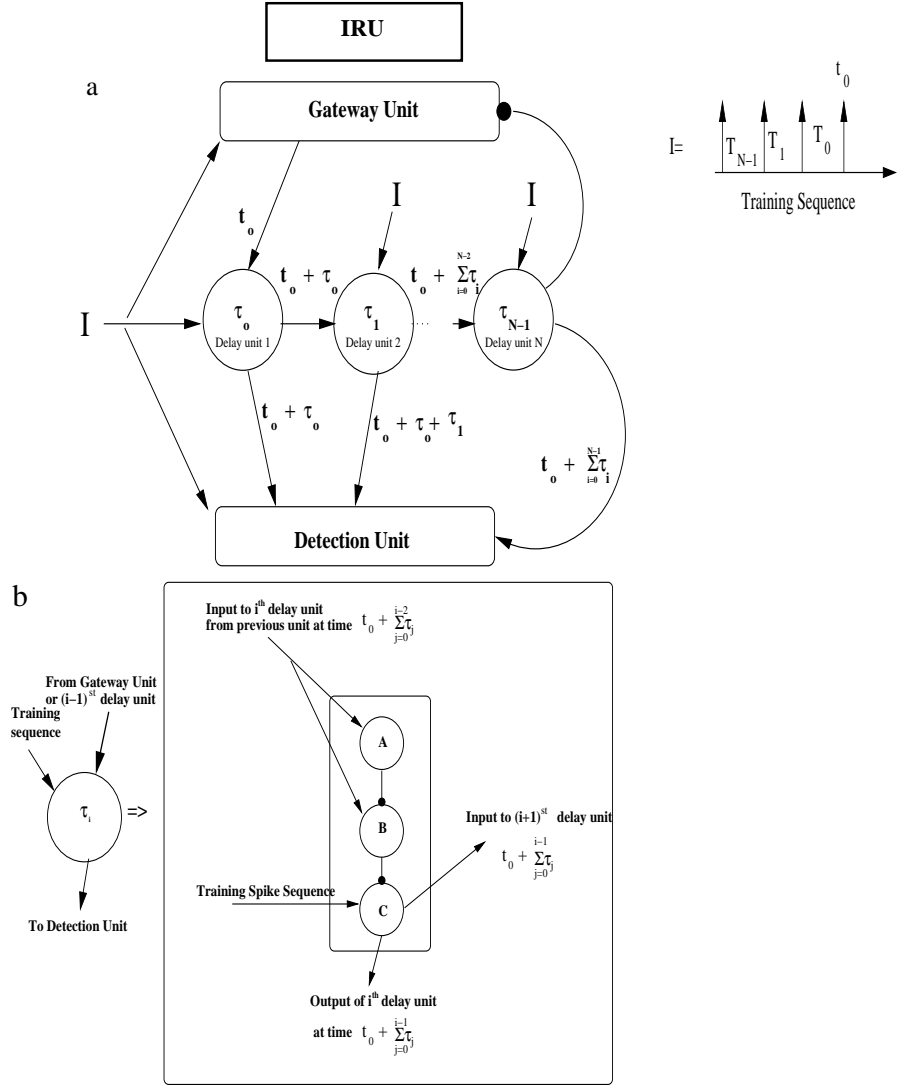}
\caption{ (a) Schematic diagram of an IRU. It is comprised of
three parts, a gateway unit, a subcircuit of time delay units, and
a detection unit. The gateway unit synchronizes the phase of the
input ISI sequence with the action of the rest of the IRU and
permits the first spike in the sequence to pass through and
stimulate the sequence of time delay units. The gateway unit is
turned off by inhibition until the last time delay units acts to
restore it to its resting state. The time delay units create a
replica of the ISI sequence, and the detection unit fires when two
spikes within a small window, here 1 ms, are presented to it. (b)
The abbreviated graphic for an individual time delay in the IRU
shown in Figure 6a. The neurons A and B of each delay unit receive
input from the previous delay unit. Depending on the strength of
the inhibitory synapse from B$\rightarrow$C, a spike is produced
by neuron C, after a certain delay. This is then sent to detection
unit for comparison to the original ISI input and to the next
delay unit to initiate its activity. \label{IRU}}
\end{center}
\end{figure*}

\begin{figure*}[ht!]
\includegraphics[width=5in,scale=1,angle=0]{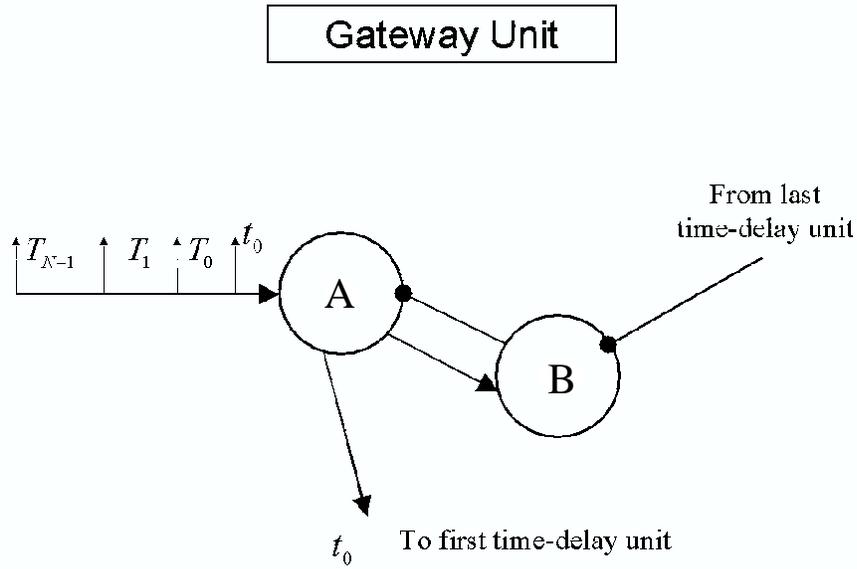}
\caption{Schematic diagram of the gateway unit for spikes entering
the IRU. The function of this unit is to initiate the delay
network activity by passing the first spike of input training
sequence to the delay network, such that the activity of delay
network is synchronized to that of the first spike of input
training sequence and shutting off further spikes until it is
reset by a signal from the last time delay unit in the IRU.
\label{gateway}}
\end{figure*}

\begin{figure*}[ht!]
\begin{center}
\includegraphics[width=7in,scale=1,angle=0]{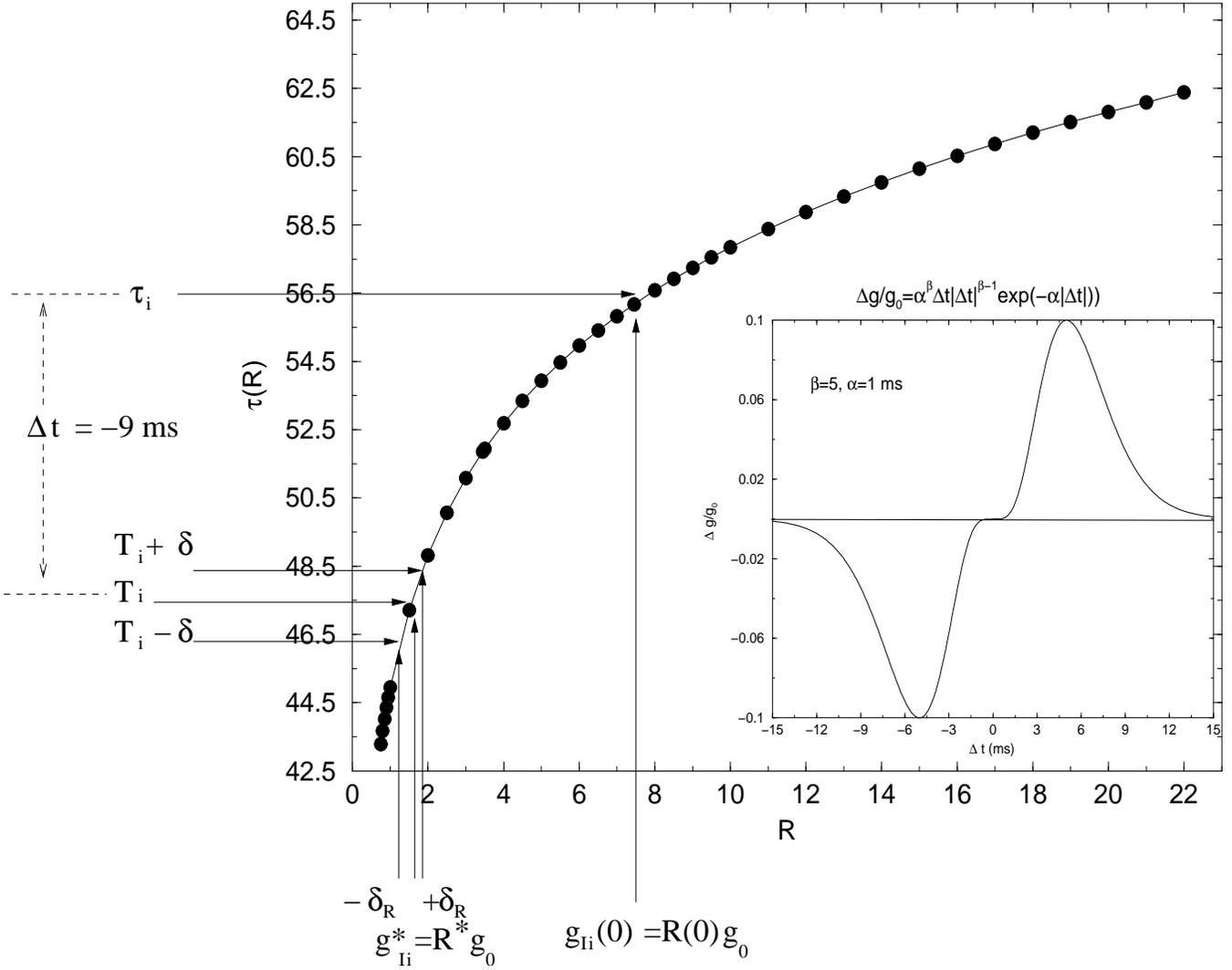}
\caption{ IRU learning: This is an example showing how the
inhibitory synaptic mechanism plays a role in modulating the
synaptic strength of the B$\rightarrow$C synapse of a given delay
unit in the IRU. The insert shows the inhibitory plasticity rule
we use in this paper~\cite{Haas}. \label{learnrule}}
\end{center}
\end{figure*}

\begin{figure*}[ht!]
\begin{center}
\includegraphics[width=5.50in,scale=1,angle=-90]{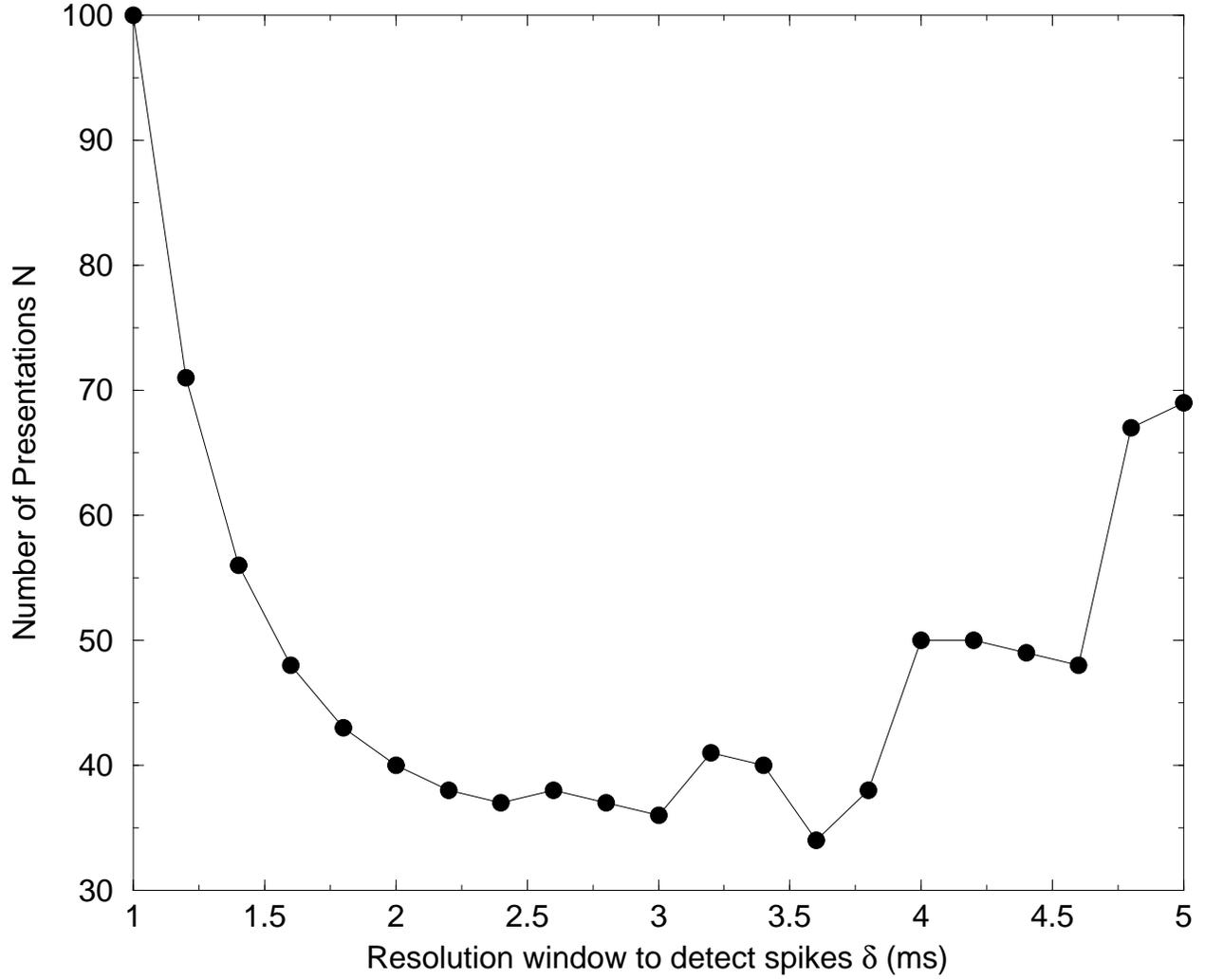}
\caption{ Convergence rate of learning of IRU, as function of the
resolution scale $\delta$ for Detection unit to recognise ISI in
the incoming spike sequence is plotted for a particular case of
noise free ISI sequence of 46,51.7,56.5 and 61.7 ms , with initial
delays for each delay unit being set at $\tau_{i}(0)=52.68$
\label{Converge}}
\end{center}
\end{figure*}

\begin{figure*}[ht!]
\begin{center}
\includegraphics[width=4.in,scale=1,angle=0]{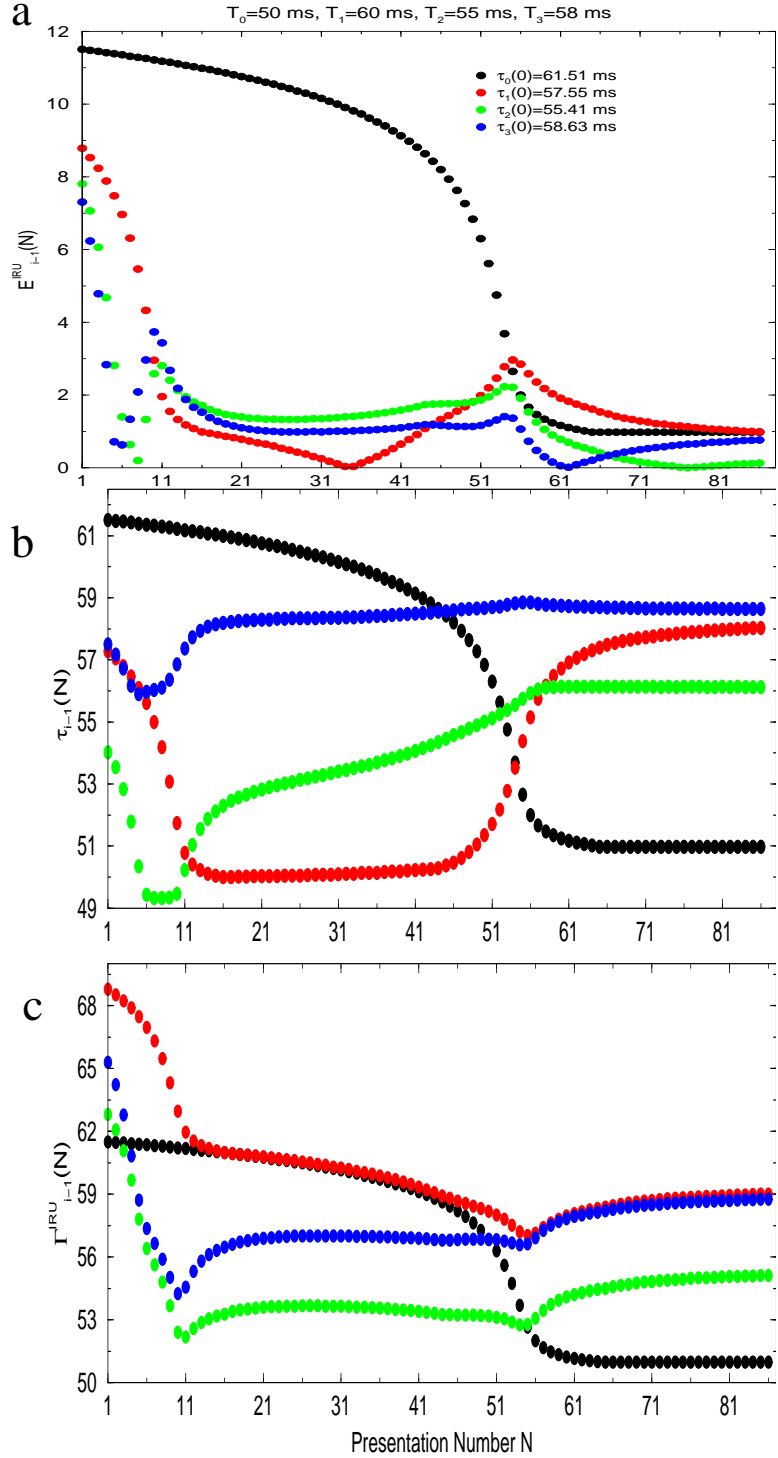}
\caption{Training an IRU to a desired ISI sequence consisting of 4
ISIs at $50,60,55,$ and $58$ ms.(a) The error in detection of ISIs
as function of trial presentation number. The training continues
until the error in detection of each ISI is within 1 ms, the
resolution threshold for the detection unit. (b) The actual delay
produced by each delay unit as learning modulates the inhibitory
synaptic strength of the B$\rightarrow$C connection in each delay
unit is plotted as function of training number N. (c) The ISIs
produced by each delay unit of the IRU. \label{train1}}
\end{center}
\end{figure*}

\begin{figure*}[ht!]
\begin{center}
\includegraphics[width=7.5in,scale=1,angle=0]{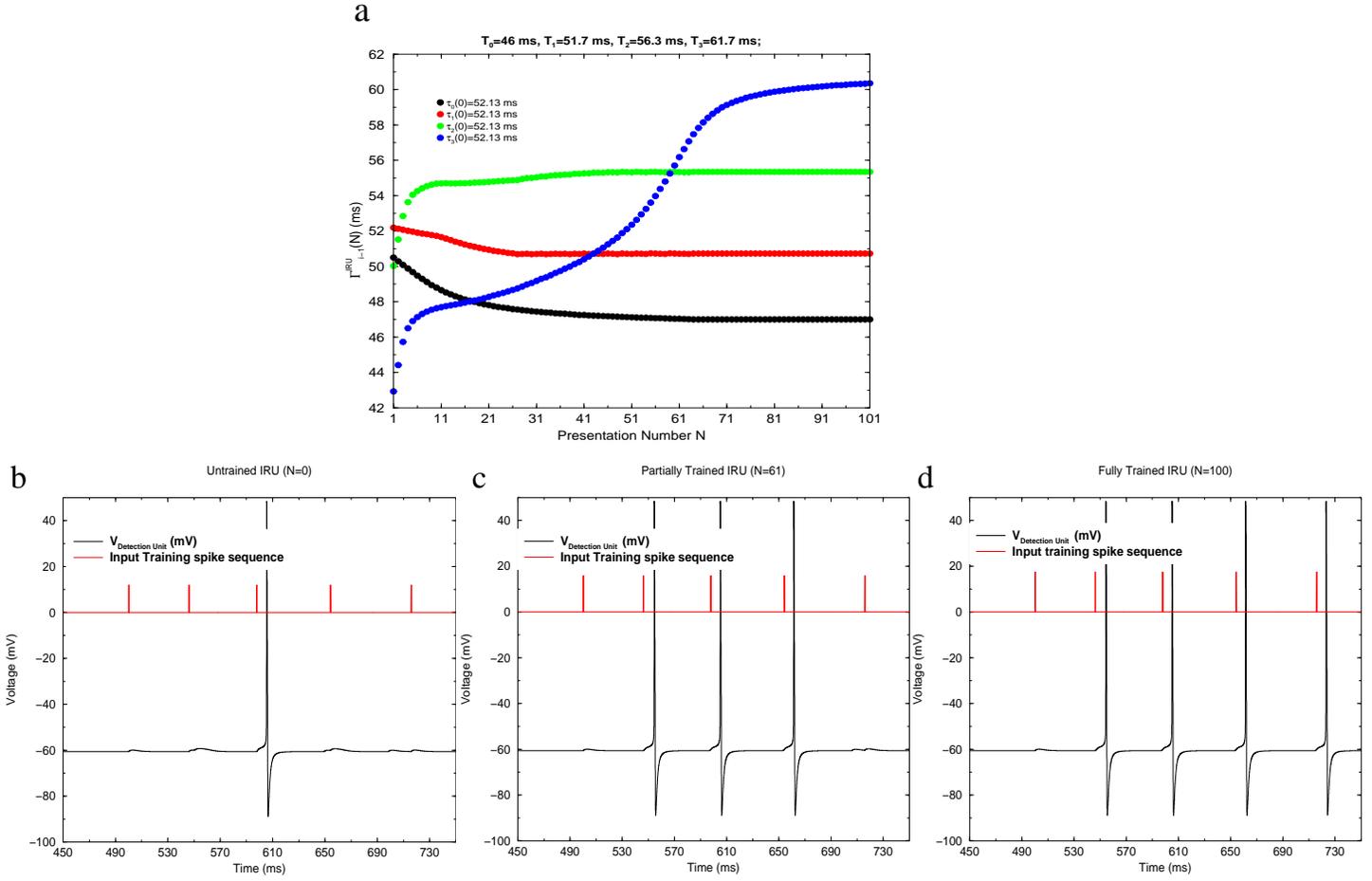}
\caption{(a) Training of an IRU on the ISI sequence, $46, 51.7,
56.3, 61.7$ ms when all time delays are initially (N=0) set to
52.13 ms. (b) Output of the detection unit when training begins.
For the selected initial conditions, the detection unit is able to
detect the 2nd ISI correctly at the beginning of training. This is
a coincidence of our choice of initial conditions. The other ISIs
are not yet trained and subthreshold EPSPs are activated in the
detection unit. (c) Output of the detection unit in the middle of
training. At this stage three of the ISIs in the selected ISI
sequence have been matched by the IRU. (d) Output of the detection
unit at the end of training when all ISIs in the input sequence
have been matched in the IRU. In each of figures 8a, 8b and 8c,
the red trace is the input training spike sequence.
\label{train2}}
\end{center}
\end{figure*}

\begin{figure*}[ht!]
\begin{center}
\includegraphics[width=5.50in,scale=1,angle=0]{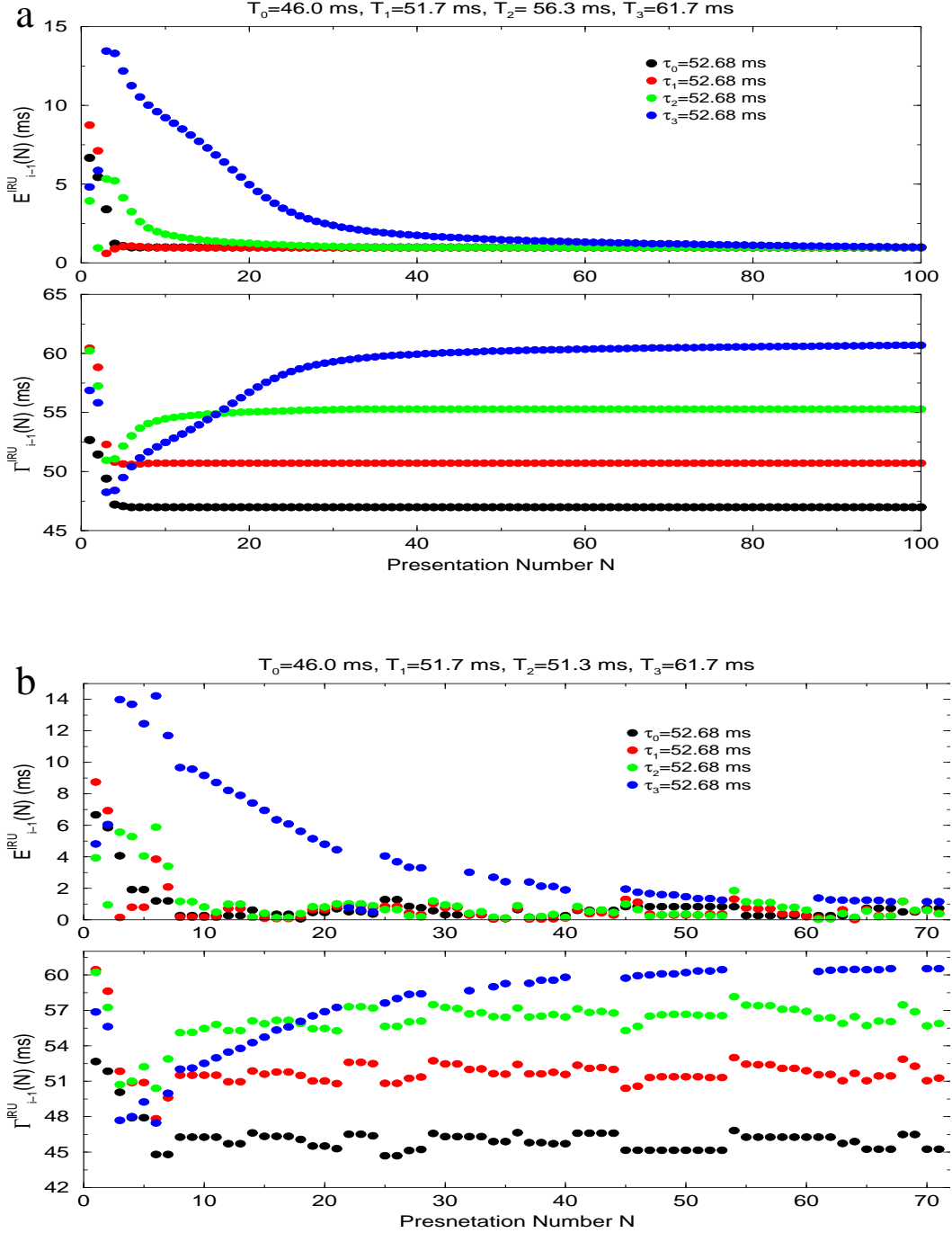}
\caption{(a) {\bf Deterministic ISIs}. In the top panel, the error
in the detection of the input ISI sequence for each delay unit is
shown. In the bottom panel, we show how the delays of each time
delay unit are modified with training to detect the ISI sequence
consisting of 4 ISIs: $46.0,51.7,56.3$ and $61.7$ ms. (b) {\bf
ISIs with jitter}. We present the same ISI sequence, but with a
jitter of $\pm\,$2 ms around the mean ISIs. In the top panel we
again present the evolution of error in ISI detection in each
delay unit. We see that it, in fact, takes fewer training sequence
iterations than in the jitter free case, and, overall, the
learning in the presence of this level of noise proceeds nearly as
efficiently as in the deterministic scenario of Figure 9a. In the
bottom panel the evolution in the presence of jitter of time
delays from the delay units with learning is presented.
\label{Noise1}}
\end{center}
\end{figure*}

\begin{figure*}[ht!]
\begin{center}
\includegraphics[width=5.50in,scale=1,angle=0]{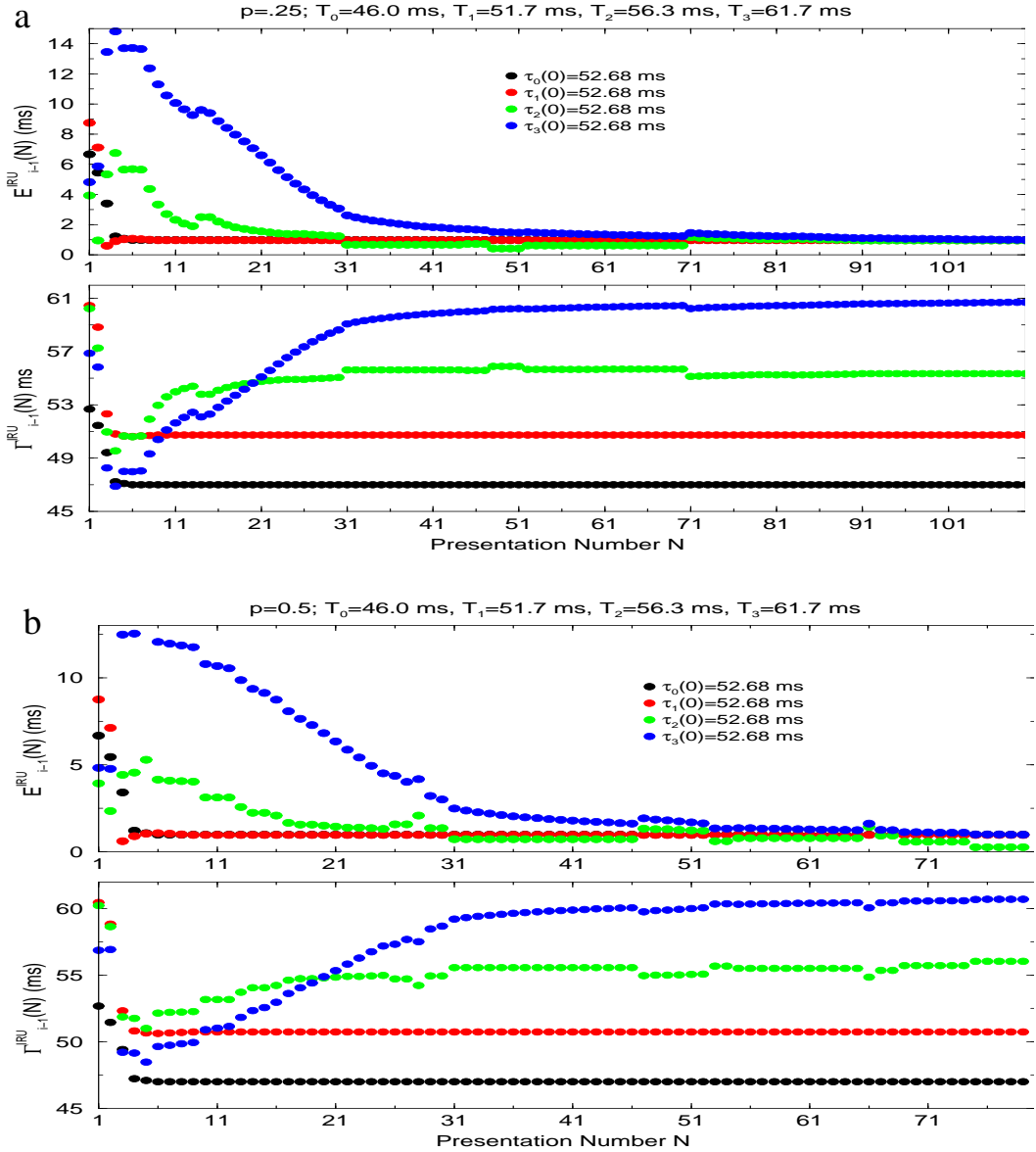}
\caption{ (a) Training the IRU on the deterministic ISI sequence
46, 51.7, 56.3, 61.7 ms in the presence of a random additional
spike in the 3rd ISI with probability p= 0.25 (b) Training the IRU
on the deterministic ISI sequence  46, 51.7, 56.3, 61.7 ms in the
presence of a random additional spike in the 3rd ISI, with
probability p= 0.5. In each case, the initial delays for each time
delay unit (N=0) are  set to 52.16 ms. \label{Noise2}}
\end{center}
\end{figure*}

\begin{figure*}[ht!]
\begin{center}
\includegraphics[width=5.5in,scale=1,angle=-90]{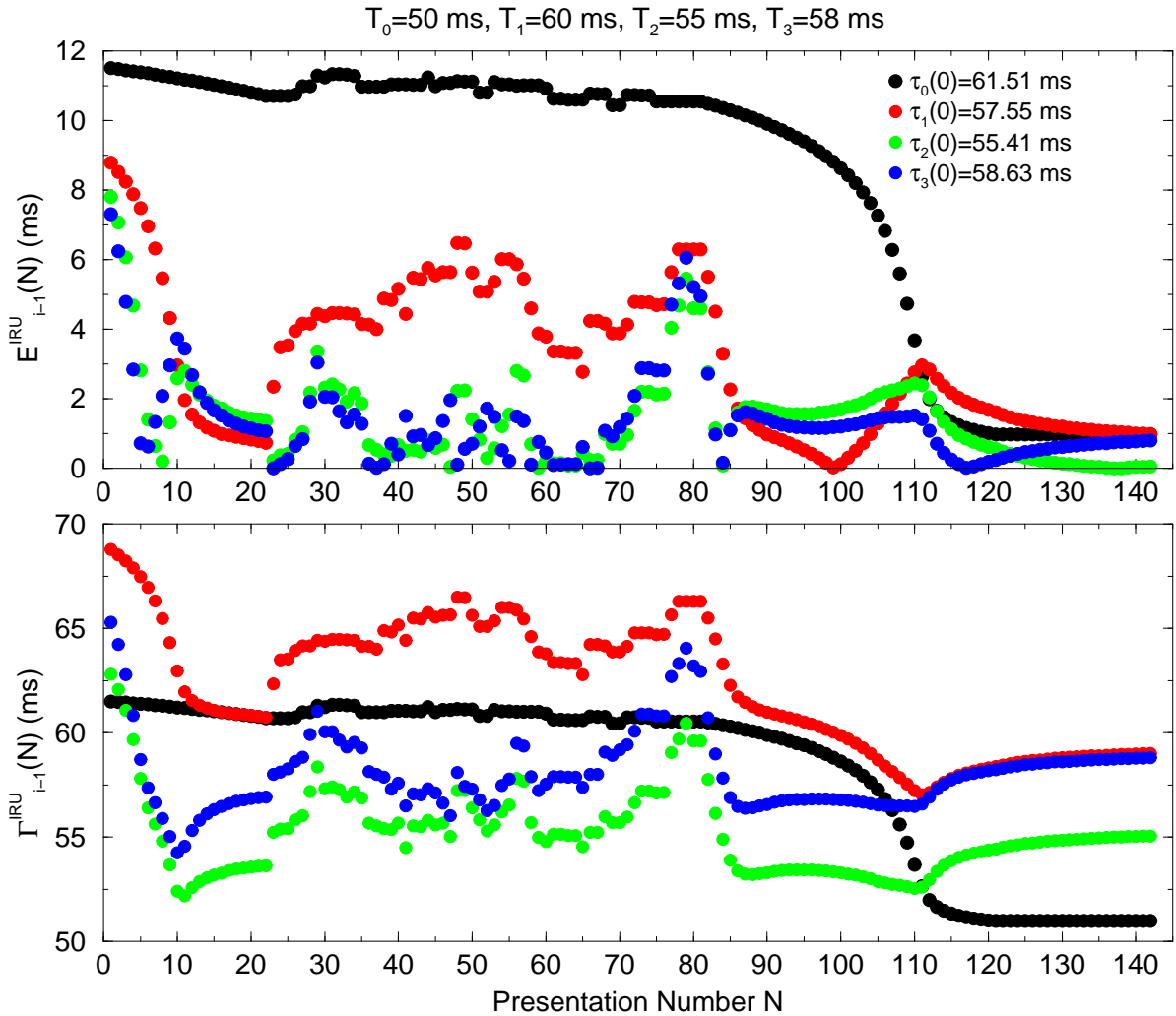}
\caption{An IRU is initially trained on the sequence of ISIs: 50
ms, 60 ms, 55 ms, and 58 ms (as in Figure 7) for 20 presentations
of the ISI sequence. Then the original sequence is replaced by an
ISI sequence with the same mean values of the ISIs but with
normally distributed noise of RMS variation 5 ms for 60
presentations. After this we make a further eighty presentations
of the original noise free ISI sequence. As a conjecture we might
connect the degeneration of the trained IRU in the presence of a
noisy ISI sequence to the degeneration of song when a bird is
deafened and no longer receives precise ISI sequences representing
its own song at NIf or HVC. \label{Noise3}}
\end{center}
\end{figure*}

\begin{figure*}[ht!]
\begin{center}
\includegraphics[width=7in,scale=1,angle=0]{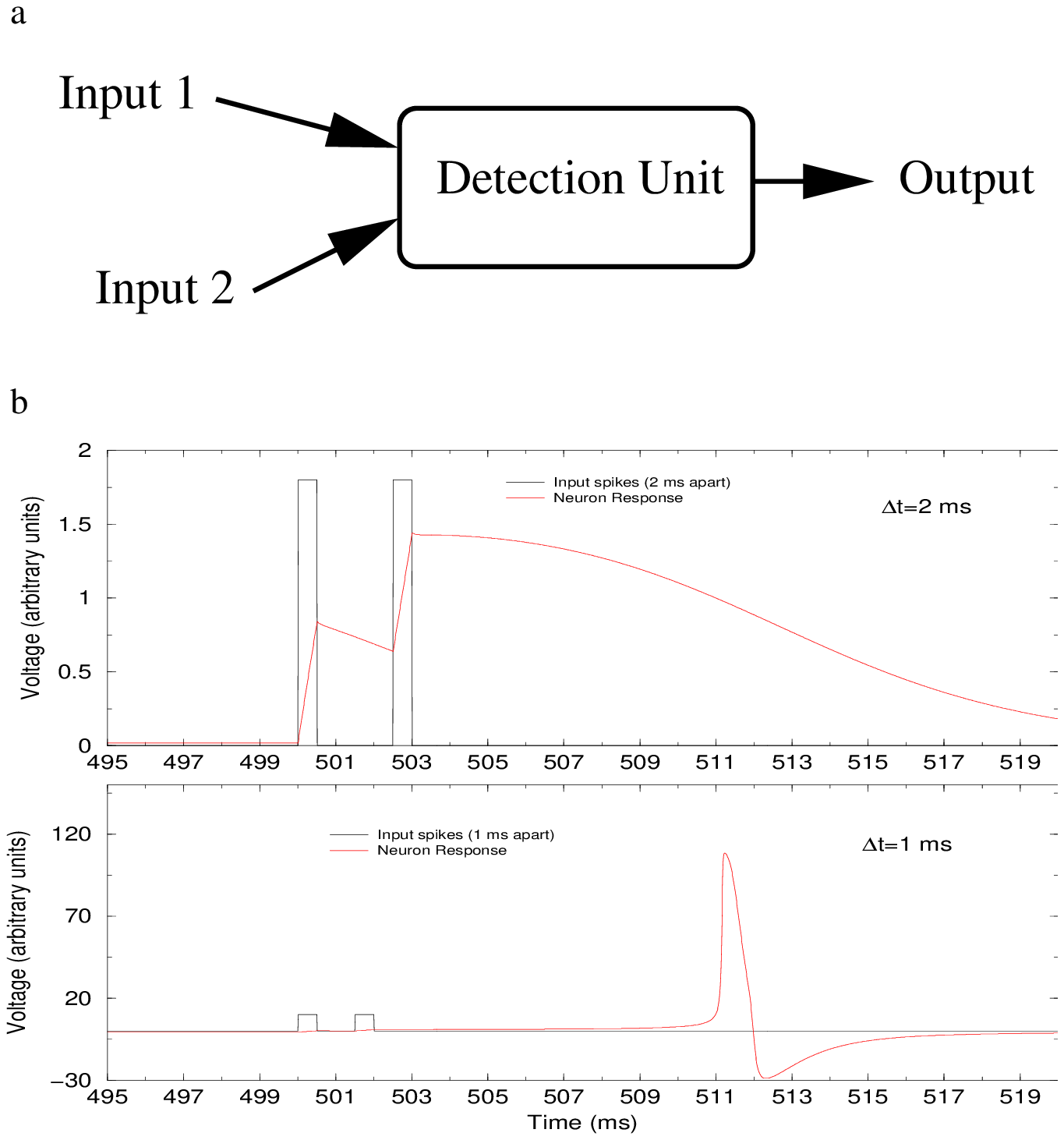}
\caption{(a) Schematic of the detection unit. It receives two
input spikes at various time delays. It responds with a spike if
the two inputs are within 1 ms of each other. (b) {\bf Top panel}
The scaled response of the detection unit when two input arrive
within 2 ms of each other. We see that the integrated input
arriving at this delay does not result in neuron spiking. In the
{\bf bottom panel} we show the scaled neuron response to two input
spikes arriving within 1 ms of each other. The detection unit
produces a spike output, indicating coincidence detection.
\label{Detect}}
\end{center}
\end{figure*}

\begin{figure*}[ht!]
\begin{center}
\includegraphics[width=7in,scale=1,angle=0]{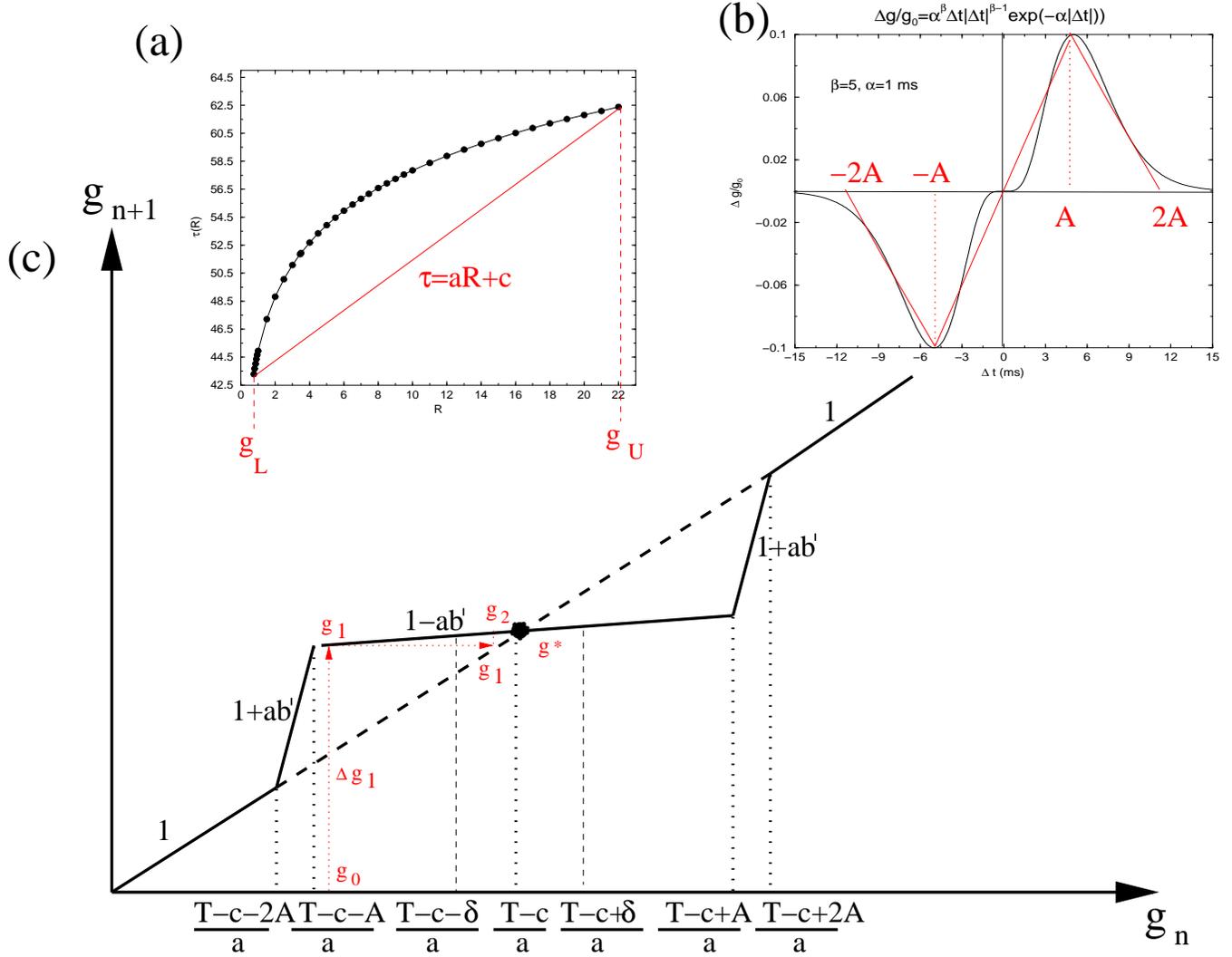}
\caption{(a) Linear approximation for the delay produced by
individual delay unit as function of inhibitory synaptic strength
(b) Linear approximation of the learning rule observed for
inhibitory synapse in entorhinal cortex. (c) The linearmap for
evolution of inhibitory synaptic strength is depicted. Sample
trajectory for evolution of the inhibitory synaptic strength
following the linear learning rule is shown in red.
\label{LinearRule}}
\end{center}
\end{figure*}

\end{document}